\documentclass[aps,prd,twocolumn,showpacs,superscriptaddress,groupedaddress,nofootinbib]{revtex4}  

\usepackage{graphicx}
\usepackage{dcolumn}
\usepackage{bm}

\usepackage{graphicx}	
\usepackage{amsmath}	
\usepackage{amssymb}	

\usepackage{epsfig,amsmath,natbib}
\usepackage{color,subfigure}
\usepackage{xcolor}

\usepackage{mathrsfs,amssymb,amstext}
\usepackage{ulem}
\usepackage{url}
\usepackage{bm}
\usepackage{adjustbox}

\begin{document}

\preprint{APS/123-QED}

\title{Cosmic Radio Background from Primordial Black Holes at Cosmic Dawn}



\author{Zhihe Zhang}
\affiliation{Department of Physics, College of Science, Tibet University, Lhasa 850000, P. R. China}  
\affiliation{Key Laboratory of Cosmic Rays (Tibet University), Ministry of Education, Lhasa 850000, P. R. China} 
\affiliation{National Astronomical Observatories, Chinese Academy of Sciences, 20A Datun Road, Chaoyang District, Beijing 100101, P. R. China} 
\affiliation{School of Astronomy and Space Science, University of Chinese Academy of Sciences, Beijing 100049, P. R. China}

\author{Bin Yue}%
\email{yuebin@nao.cas.cn}
\affiliation{National Astronomical Observatories, Chinese Academy of Sciences, 20A Datun Road, Chaoyang District, Beijing 100101, P. R. China}
 
\author{Yidong Xu}
\affiliation{National Astronomical Observatories, Chinese Academy of Sciences, 20A Datun Road, Chaoyang District, Beijing 100101, P. R. China}

\author{Yin-Zhe Ma}
\affiliation{School of Chemistry and Physics, University of KwaZulu-Natal, Westville Campus, Private Bag X54001, Durban 4000, South Africa}
\affiliation{NAOC-UKZN Computational Astrophysics Centre (NUCAC), University of KwaZulu-Natal, Durban, 4000, South Africa}
\affiliation{National Institute for Theoretical and Computational Sciences (NITheCS), South Africa}

\author{Xuelei Chen}
\affiliation{National Astronomical Observatories, Chinese Academy of Sciences, 20A Datun Road, Chaoyang District, Beijing 100101, P. R. China}
\affiliation{School of Astronomy and Space Science, University of Chinese Academy of Sciences, Beijing 100049, P. R. China}
\affiliation{Department of Physics, College of Sciences, Northeastern University, Shenyang, 110819, China}
\affiliation{Center of High Energy Physics, Peking University, Beijing 100871, China}

\author{Maoyuan Liu}
\email{liumy@utibet.edu.cn}
\affiliation{Department of Physics, College of Science, Tibet University, Lhasa 850000, P. R. China}  
\affiliation{Key Laboratory of Cosmic Rays (Tibet University), Ministry of Education, Lhasa 850000, P. R. China}

\date{\today}

\begin{abstract}
The presence of an extra radio background besides the cosmic microwave background has important implications for the observation of the 21-cm signal during the cosmic Dark Ages, Cosmic Dawn, and epoch of Reionization. The 
strong absorption trough found in the 21-cm global spectrum measured by the EDGES experiment, which has a much greater depth than the standard model prediction, has drawn great interest to this scenario, but more generally it is still of great interest to consider such a cosmic radio background (CRB) in the early Universe.  To be effective in affecting the 21-cm signal at early time, such a radio background must be produced by sources which can emit strong radio signals but modest amount of X-rays, so that the gas is not heated up too early. We investigate the scenario that such a radio background is produced by the primordial black holes (PBHs). For PBH with a single mass, we find that if the PBHs' abundance $\log(f_{\rm PBH})$ (ratio of total PBH mass density to total matter density) and mass satisfy the relation $\log(f_{\rm PBH}) \sim -1.8\log(M_\bullet/{\rm M}_{\odot})-3.5$ for $1\,{\rm M}_\odot \lesssim M_\bullet \lesssim 300 {\rm M}_\odot$, and have jet emission, they can generate a CRB required for reproducing the 21-cm absorption signal seen by the EDGES. The accretion rate can be boosted if the PBHs are surrounded by dark matter halos, which permits lower $f_{\rm PBH}$ value to satisfy the EDGES observation. In the latter scenario, since the accretion rate can evolve rapidly during the Cosmic Dawn, the frequency (redshift) and depth of the absorption trough can determine the mass and abundance of the PBHs simultaneously. For absorption trough redshift $\sim$ 17 and depth $\sim -500$ mK, it corresponds to $M_\bullet \sim 1.05\,{\rm M}_{\odot}$ and $f_{\rm PBH}\sim 1.5\times10^{-4}$.

\end{abstract}

\maketitle


\section{Introduction}\label{sec:intro}

The 21-cm global spectrum is the result of the thermal evolution of the intergalactic medium (IGM), and cosmic star formation history during the Dark Ages, Cosmic Dawn, and Reionization  
\cite{21cm_review1_2006PhR...433..181F,21cm_review2_2012RPPh...75h6901P}.
Standard cosmology models typically predicts an absorption trough at Cosmic Dawn, with depth up to $\sim 200$ mK \cite{Chen2004ApJ,Chen2008ApJ}.  
The location, depth, and detailed shape of the absorption trough reflect complicated interactions between the IGM hydrogen atoms, the X-ray and Ly$\alpha$ photons from the first stars and galaxies, and the cosmic microwave background (CMB) photons from the early Universe.

There have been several ground-based single-antenna experiments to measure the 21-cm global spectrum, such as EDGES  
\cite{21cm_moreCRB_1810.05912},
SARAS  
\cite{experiment_SARAS2_1703.06647},
SCI-21~\cite{experiment_SCI-21_1311.0014},  LEDA~\cite{experiment_LEDA_1709.09313}, BIGHORNS~\cite{experiment_BIGHORNS_1509.06125},   
REACH   
\cite{REACH_2022NatAs}, etc.
There are also proposals to carry out the measurements in space,  such as the DARE/DAPPER~\cite{future_mission_DAPPER_2003.06881}, DSL~\cite{future_emission_DSL_2007.15794},  and FARSIDE~\cite{future_mission_FARSIDE_2103.08623} projects. The EDGES experiment detected an absorption trough on 21-cm global spectrum~\cite{21cm_moreCRB_1810.05912}. The frequency is in agreement with the 21-cm signal from the era of Cosmic Dawn ($z\sim17$)~\cite{21cm_moreCRB_1810.05912}, though the depth is $\sim500$ mK, much deeper than the maximum limit in the standard cosmology model~\cite{Xu2018ApJ,EDGES_stronger_2102.12865}. The subsequent SARAS-3 experiment obtained a result contradictory with that of the   EDGES~\cite{SARAS3_2112.06778}, but it may be still too early to reach a final conclusion ~\cite{Chen2022NatAs}, as the experiment systematics are far from understood. In this paper we will take the view that the EDGES result represents a possible 21-cm global spectrum which requires a theoretical explanation.

Several theoretical models have been proposed to explain the surprising result of the EDGES experiment. These include the exotic dark matter particles which couple to and cool the baryonic gas~\cite{Barkana2018Natur}, and the extra cosmic radio background (CRB) in addition to the CMB~\cite{21cm_moreCRB_1802.07432,BH_CRB_1803.01815}. Although the existence of this deep absorption signal is still debated, the scenario that at Cosmic Dawn there is an extra radio background and its impact on the 21-cm global spectrum is nevertheless an interesting and novel perspective.
Many objects may produce an extra CRB at Cosmic Dawn, for example the first galaxies in small halos with high efficiency of star formation and radio production~\cite{Mirocha2019MNRAS},
dark matter annihilations~\cite{Yang_2018}, the first black holes~\cite{BH_CRB_1803.01815,Mirabel2022NewAR}, and the primordial black holes (PBHs).

PBHs may arise from small-scale primordial curvature fluctuations, and form in the early Universe, much earlier than the formation of the first generation stars (Pop III stars)~\cite{PBH_CRB_2108.13256,Cang2021,Cang2022}. In addition, they can be much more abundant than the black hole seeds formed after the death of Pop III stars. Therefore PBHs have the potential to generate a significant amount of the CRB through its accretion and jet mechanisms. In most models, the initial PBH mass ranges from the Planck mass to $\sim10^5{\rm M}_{\odot}$~\cite{planck_hawking_1,upperlimit_0412134}. PBHs with mass $M_{\bullet}<10^{15}$ g would have evaporated through Hawking radiation~\cite{Carr1979,PBHreview5}. The more massive ones could however survive and grow by accreting the ambient medium~\cite{upperlimit_0412134}. They may form accretion disks with a significant angular momentum~\cite{CMB_bound_disk_PBH_1707.04206}, or even launch relativistic jets like astrophysical black holes  
\cite{PBH_syn_2111.08699}.
So far no PBH has been detected, but the abundance has been constrained by several observations, such as the gravitational wave 
\cite{GW_PBH_1603.00464},
gravitational lensing~\cite{lensing_PBH_1204.2056}, the impact of accretion on the cosmic thermal evolution~\cite{CMB_bound_disk_PBH_1707.04206}, CMB spectra distortion~\cite{CMBbounds_Serpico_2002.10771,PBHreview_2007.10722} and CMB anisotropy power spectrum~\cite{Cang2021,Cang2022}.

The CRB from PBHs has been investigated by many authors.   
Most of these works focus on the Hawking radiation of smaller PBHs~\cite{PBH_CRB_2107.02190,PBH_CRB_2108.13256,Saha2022PhRvD,Mittal2022MNRAS}, or the free-free emission from accretion disks of stellar-mass PBHs~\cite{Tashiro2021_ff,Y1612}.
In these scenarios, it is not easy for the PBHs to produce sufficient CRB to boost the 21-cm absorption significantly, otherwise the IGM would be ionized/heated too early, in conflict with the observational constraints on the reionization history~\cite{Acharya2022MNRAS}.
It is generally unquestionable that the PBH can produce some X-rays \cite{Mena2019PhRvD}. If there is also radio emission associated with the X-ray radiation, then PBHs can produce a sufficient amount of CRB and boost the 21 cm absorption to the EDGES level \cite{Hasinger2020JCAP}.
Since these PBHs also contribute to building up the Cosmic X-ray Background (CXB),
the observed CXB 
also put tight limits on the contribution to CRB from PBHs~\cite{Ziparo2022MNRAS}.

Astrophysical black holes can usually launch powerful jet which produces strong synchrotron radiation~\cite{Jet3_1711_01292,jet_2108.12380}. The same effect may also exist for PBH. PBH usually sits in the gaseous environment of the interstellar medium, so its accretion disk can naturally launch outflows such as winds and jets. Ref.~\cite{PBH_syn_2111.08699} showed that PBHs with significant spin can sustain powerful relativistic jets and generate associated cocoons. Ref. \cite{Hasinger2020JCAP} pointed out that the PBH can play important role in amplifying the initial magnetic field. If the jet has a strong magnetic field, then its synchrotron could dominate its radio radiation. As a result, the synchrotron from PBHs may generate a much stronger CRB than the free-free emission, and influence the 21-cm absorption significantly. This is the basic motivation of our work. In this paper, we investigate if in this case PBHs can produce a strong CRB that boosts the 21-cm absorption signal at the Cosmic Dawn, while at the same time only heats the IGM slightly.

The paper is organized as follows. In Sec.~\ref{sec:methods} we consider the accretion rate evolution of the PBHs. We estimate the contribution to the CRB, the thermal history of the IGM, and the influence on the 21-cm signal at Cosmic Dawn, if an empirical radio luminosity$-$X-ray luminosity relation is applicable to the PBHs.  In Sec.~\ref{sec:jet} we calculate the synchrotron radiation of PBHs using a jet model and evaluate the final influence on the 21-cm signal. Finally, we summarize the results and discuss some issues of this problem in Sec.~\ref{sec:summary}.

\section{Methods}\label{sec:methods}
\subsection{The accretion rate and radiative efficiency of PBHs}
\subsubsection{Naked PBHs}

First we consider a naked PBH, by which we mean one without a surrounding dark matter halo.
The accretion process of a massive PBH has been studied comprehensively in the past years (e.g. \citep{Ricotti2007_lambda, Ricotti2007_macc,Y1612,Tashiro2021_ff}). 
Ignoring the angular momentum of the accreted gas, the accretion is described well by the spherical symmetric model and the accretion rate is~\cite{Bondi1952}
\begin{equation}
\dot{M}_{\rm B}=4 \pi \lambda \bar{\rho}_{\rm b} v_{\rm eff} r^2_{\rm B},
\label{eq:M_dot}
\end{equation}
where the $v_{\rm eff}=\sqrt{v_{\rm rel}^2+c_{\rm s}^2}$, $v_{\rm rel}$ is the velocity of the relative motion between the black hole and the ambient gas, and $c_{\rm s}$ is the sound speed.
$\lambda$ is a dimensionless parameter describing the net accretion; and $\bar{\rho}_{\rm b}=\Omega_{\rm b}\rho_{\rm cr}(1+z)^3$ is the cosmic mean baryon density. $\rho_{\rm cr}=2.775\,h^{2}\times 10^{11}\,{\rm M}_{\odot}\,{\rm Mpc}^{-3}$ is the critical density of the present Universe.
The Bondi radius of a PBH with mass $M_{\bullet}$ is defined as the radius for which the escape velocity equals $v_{\rm eff}$:
\begin{equation}
r_{\rm B}=\frac{GM_{\bullet}}{v_{\rm eff}^2}. \label{eq:rB}
\end{equation}

The dark matter may have a relative motion with respect to the gas, once the two are both decoupled from each other and from the photons~\cite{Ricotti2007_macc}.
A PBH is therefore expected to  have a proper motion relative to its ambient gas medium. In linear regime, the dispersion of the relative velocity  $\langle v_{\rm L}^2 \rangle ^{1/2}\approx {\rm min}[1,z/10^3]\times30~{\rm km/s}$, and the effective velocity is \cite{Y1612},
\begin{equation}
v_{\rm eff}\approx \begin{cases} 
\sqrt{c_{\rm s}\langle v_{\rm L}^2 \rangle ^{1/2}}& \langle v_{\rm L}^2 \rangle ^{1/2} \gg c_{\rm s} \\
c_{\rm s}&   \langle v_{\rm L}^2 \rangle ^{1/2} \ll c_{\rm s}.
\end{cases}
\label{eq:v_eff}
\end{equation}
The relative motion reduces the accretion rate, since $\dot{M}_{\rm B}\propto v_{\rm eff}^{-3}$ .

The sound speed $c_{\rm s}$ depends on the kinetic temperature of the IGM gas, $T_{\rm k}$, while the gas is also heated by the X-ray radiation from PBHs, so $T_{\rm k}$ must be solved self-consistently from the cosmic thermal evolution equations with X-ray heating from PBHs.

The parameter $\lambda$ is determined by the PBH surrounding environment as well as various physical processes such as Hubble expansion, Compton drag and cooling of the CMB photons, and negative feedback. 
We calculate it following the Refs.~\cite{Ricotti2007_lambda,Y1612}. 
For a naked PBH, 
\begin{eqnarray}
\lambda(\gamma,\beta) &=& \frac{1}{\lambda_{\rm iso}}\left[\lambda_{\rm ad}+(\lambda_{\rm iso}-\lambda_{\rm ad})\left(\frac{\gamma^{2}_{\rm c}}{88+\gamma^{2}_{\rm c}}\right)^{0.22}\right] \nonumber \\
&\times & \left[\frac{1}{(\sqrt{1+\beta_{\rm c}}+1)^2}\exp\left(\frac{9/2}{3+\beta_{\rm c}^{3/4}}\right)\right],
\label{eq:lambda_nohalo}
\end{eqnarray}
where $\lambda_{\rm ad}=(3/5)^{3/2}/4\approx0.12,\lambda_{\rm iso}=e^{3/2}/4\approx1.12$;
$\beta_{\rm c}$ and $\gamma_{\rm c}$ are the dimensionless Compton drag and cooling rates from the surrounding CMB photons respectively~\cite{Y1612}:
\begin{eqnarray}
    \beta_{\rm c} &=&\frac{4}{3} \frac{x_{\rm e}\sigma_{\rm T}\rho_{\gamma}}{m_{\rm p}c}\frac{r_{\rm B}}{v_{\rm eff}} 
    \label{eq:beta_c} \\
    \gamma_{\rm c} &=& \frac{8}{3} \frac{x_{\rm e}\sigma_{\rm T}\rho_{\gamma}}{m_{\rm e}c(1+x_{\rm e})}\frac{r_{\rm B}}{v_{\rm eff}},
    \label{eq:gamma_c}
\end{eqnarray}
where $x_{\rm e}$ is the electron fraction in the IGM, $c$ is the speed of light, $\sigma_{\rm T}=6.65\times10^{-25}$ cm$^2$ is the Thomson scattering cross-section, and $\rho_{\gamma}=\pi^{2}T^{4}_{\rm CMB}/15$ is the energy density of the CMB.

\begin{figure}
\centering{
\subfigure{\includegraphics[width=0.45\textwidth]{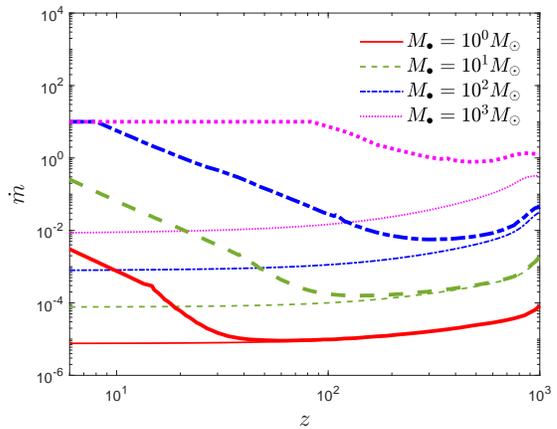}}
\caption{The evolution of the dimensionless accretion rate with redshift. The PBH masses corresponding to the curves from bottom to top are: $10^0{\rm M}_{\odot}$ (red solid), $10^1{\rm M}_{\odot}$ (green dashed), $10^2{\rm M}_{\odot}$ (blue dash-dotted), $10^3{\rm M}_{\odot}$ (magenta dotted). Of the two lines corresponding to each mass, the upper (thick) one is the case if PBHs are surrounded by dark matter halos and the lower (thin) one is the case of naked PBHs.
}
\label{fig:m_acc_tot}
}
\end{figure}

In Fig.~\ref{fig:m_acc_tot} we plot the evolution of dimensionless accretion rate $\dot{m}\equiv\dot{M}_{\rm B}/(L_{\rm Edd}/c^2)$ for naked PBHs of masses $10^0, 10^1, 10^2, 10^3 {\rm M}_{\odot}$ by thin curves, where 
$$L_{\rm Edd}=1.26\times 10^{38} M_{\bullet} ~{\rm erg~ s}^{-1}$$ 
is the Eddington luminosity. At $z\sim17$, $\dot{m}$ ranges from $\sim10^{-5}$ for $M_{\bullet}=1\,{\rm M}_{\odot}$ to $\sim10^{-2}$ for $M_{\bullet}=10^3\,{\rm M}_{\odot}$. Given these small ratios, the accretion is not affected by the Eddington limit in this case.
   
\subsubsection{PBHs inside dark matter halos}

If PBHs account for only a small fraction of the dark matter, they can attract the surrounding dark matter particles. A PBH with mass $M_{\bullet}$ will form an extended dark matter halo with mass~\cite{Ricotti2007_lambda}
\begin{equation}
    M_{\rm h}=M_{\bullet}\left(\frac{3000}{1+z}\right)
\end{equation}
and radius
\begin{eqnarray}
r_{\rm h}&=& 0.339 r_{\rm turn} \nonumber \\
&=& 0.339\frac{58}{1+z}\left(\frac{M_{\rm h}}{{\rm M}_{\odot}}\right)^{\frac{1}{3}}\text{pc},
\end{eqnarray}
where $r_{\rm turn}$ is the turnaround radius. 

We assume the dark matter halo has spherical symmetry. Because there is a black hole at the center of the dark matter halo, the interior of the density profile becomes steeper~\cite{Adamek2019PhRvD} than the predictions from the NFW model~\cite{NFW1996ApJ} or the Einasto model~\cite{Einasto1965}. Therefore, we adopt the power-law density profile derived from numerical simulations~\cite{1985ApJS_58_39B} to describe the dark matter distribution. The enclosed dark matter mass is then
\begin{equation}
    M_{\rm h}(<r)=\begin{cases}
    M_{\rm h} \times  \left(\frac{r}{r_{\rm h}}\right)^p& \text{$r<r_{\rm h}$}
    \\M_{\rm h}& \text{$r>r_{\rm h}$},
    \end{cases}
\end{equation}
where $p=0.75$. 
 
Obviously, at the redshifts of interest, the dark matter halo mass is much higher than the mass of PBH, so the velocity dispersion of dark matter is much larger than the escape velocity. Therefore, the accretion of dark matter to PBHs is negligible~\cite{Ricotti2007_macc}. For a PBH inside a dark matter halo,
basically the accretion rate is solved from conservation equations of mass and momentum, and gas thermodynamics. 
In what follows we give a brief summary of the formulae for calculating such an accretion rate. More details can be found in the original references.

Let the halo Bondi radius to halo radius ratio be $\chi \equiv r_{\rm B,{\rm h}}/r_{\rm h}$,
where $r_{\rm B,{\rm h}}= GM_{\rm h}/v_{\rm eff}^2$ is the halo Bondi radius.
When $\chi >1$, we simply replace $M_{\bullet}$ with $M_{\rm h}+M_{\bullet}$ to obtain the new accretion rate. However, when $\chi<1$, only a fraction of the halo mass plays the role of sustaining Bondi accretion. In this case one should replace the Bondi radius in Eqs. (\ref{eq:M_dot}, \ref{eq:beta_c}, \ref{eq:gamma_c}) with the effective radius solved from the following equation:
\begin{align}
    r_{\rm B,eff}&= \frac{G(M_{\bullet}+M_{\rm h}(<r_{\rm B,eff}))}{v_{\rm eff}^2} \nonumber \\
   & \approx r_{\rm B,h}\left(\frac{r_{\rm B,eff}}{r_{\rm h}}\right)^p
    \label{eq:r_Beff_h}.
\end{align}
$\lambda$ is solved from the hydrodynamic equations~\cite{Ricotti2007_macc,Y1612,Tashiro2021_ff}:
\begin{eqnarray}
-\widetilde{\rho}x^2u &=& \lambda    \label{eq:hydro_eq1} \\
-\frac{M_r}{x^2}-\frac{{\rm d}\widetilde{T}}{{\rm d}x}-\frac{\widetilde{T}}{\widetilde{\rho}}\frac{{\rm d}\widetilde{\rho}}{{\rm d}x}-\beta_{\rm c}u &=& u\frac{{\rm d}u}{{\rm d}x} \label{eq:hydro_eq2} \\
u\frac{{\rm d}\tilde{T}}{{\rm d}x}+u\widetilde{T}(-\frac{2}{3})\frac{1}{\widetilde{\rho}}\frac{{\rm d}\widetilde{\rho}}{{\rm d}x} &=& \gamma_{\rm c}(1-\widetilde{T}),
\label{eq:hydro_eq3}
\end{eqnarray}
where $x=r/r_{\rm B,eff}$ is the dimensionless radial coordinate; $u=v_r(x)/v_{\rm eff}<0$ is the dimensionless radial velocity of accreting gas; $\widetilde{\rho}=\rho_b(x)/\rho_{b,\infty}$ is the dimensionless gas density; $\widetilde{T}=T(x)/T_{\rm CMB}$ is the dimensionless temperature; and $M_r$ represents a mass proportion
\begin{equation}
    M_r=\frac{M_{\bullet}+M_{\rm h}(<r)}{M_{\bullet}+M_{\rm h}(<r_{\rm B,eff})}. 
    \label{eq:M_r}
\end{equation}

Defining $v=-u$ and replacing Eq. (\ref{eq:hydro_eq1}) with the  derivative form: 
$$ d\ln v +d\ln \widetilde{\rho} =-2d\ln x, $$
Eqs. (\ref{eq:hydro_eq1}$)-($\ref{eq:hydro_eq3}) can be transformed to the matrix form:
\begin{eqnarray}
&& \begin{pmatrix}
    v^2 & \widetilde{T} & \widetilde{T}\\
    0   & 2v/3 & -v\\
    1   & 1       & 0
    \end{pmatrix}
    \begin{pmatrix}
    {\rm d}\ln v \\ {\rm d}\ln\widetilde{\rho} \\ {\rm d}\ln\widetilde{T}
    \end{pmatrix} \nonumber \\
&& =-\begin{pmatrix}
    M_r/x-\beta_{\rm c}vx \\
    -\gamma_{\rm c}(1/\widetilde{T}-1)x \\
    2
    \end{pmatrix}
    {\rm d}\ln x.
    \label{eq:pma_hydro}
\end{eqnarray}
The boundary conditions are $\widetilde{T}(x)=1$ and $v(x)=\lambda/x^2$ at large $x$. 
There are two further requirements.
The infalling velocity $u$ must increase monotonously  towards the center; and there must exist a sonic point, where the infalling velocity is continuous. In such case a unique $\lambda$ value can be derived from Eq. (\ref{eq:pma_hydro}). Details can be found in Ref.~\cite{Y1612}.

When $M_r=1$, the hydrodynamic equation (\ref{eq:pma_hydro}) reduces to the naked PBH model. We have checked that the solutions are indeed consistent with Eq. \eqref{eq:lambda_nohalo}.

In Fig.~\ref{fig:m_acc_tot} the dimensionless accretion rate $\dot{m}$ for PBHs inside dark matter halos is shown by the thick curves.  
When the PBH is surrounded by a dark matter halo, the larger $M_{\bullet}$, the earlier the dark matter halo plays a significant role in enhancing the accretion rate. The accretion rate is boosted by $\gtrsim 2$ orders of magnitude at $z\sim17$. For PBHs of $M_{\bullet}\gtrsim10^{2}\,{\rm M}_{\odot}$, the accretion rate can reach the Eddington limit of the central PBH before $z\sim10$.

\subsubsection{The bolometric luminosity}

For a PBH with an accretion rate $\dot{m}$, the bolometric luminosity can be written as
\begin{equation}
L_{\rm bol}=\eta \dot{m} L_{\rm Edd},
\end{equation}
where $\eta$  is the radiative efficiency and it depends on $\dot{m}$ and the geometry of the accretion flow. 
The geometry also depends on the accretion rate.
If $\dot{m} \gtrsim 1$, then it forms the standard thin accretion disc\footnote{In some literatures $\dot{m}$ is defined as $\dot{M}_{\rm B}/(10L_{\rm Edd}/c^2)$, using this definition the critical accretion rate for  accretion mode transition is $\sim0.1$.}; if $\dot{m}\lesssim1$  then an advection dominated accretion flow (ADAF) forms~\cite{ADAF_1401.0586,jet_2108.12380,PBH_CRB_1803.09697}. For the ADAF accretion mode, when $\dot{m}\lesssim0.1$, the radiative efficiency is low and proportional to the accretion rate; when $\dot{m}\gtrsim0.1$, the radiative efficiency approaches the thin disc which has a constant radiative efficiency. Therefore \cite{Narayan2008NewAR}
\begin{equation}
\eta=0.1\times \begin{cases}
\frac{\dot{m}}{0.1} ~~&\dot{m} \le 0.1\\
1~~&\dot{m} > 0.1.
\end{cases}
\end{equation}
Since $L_{\rm bol} \le L_{\rm Edd}$, we set a hard limit on $\dot{m}$ so that $\eta \dot{m}\le 1$ always holds.

The typical value of $\lambda$ calculated from Eqs. (\ref{eq:lambda_nohalo}) or (\ref{eq:pma_hydro}) is $\mathcal{O}(0.1)$. Note however that when the accretion is non-spherical, $\lambda$ can be an order of magnitude smaller \cite{Xie2012}. If $\lambda$ is smaller, we will need a larger PBH mass to produce the same amount of radiation.

\subsection{An
empirical $L_{\rm R}-L_{\rm X}$ relation
}

Astrophysical black holes usually have strong X-ray signal that is either from the hot accretion flow (ADAF), or from the standard thin disc and the hot corona above the disc. If the black hole has a jet, then it also produces strong radio signal by the synchrotron emission of relativistic electrons inside the jet. Jet can also produce strong X-ray signal and it may dominate the X-ray radiation of the whole black hole when the accretion rate is extremely low \cite{Yuan2005}.

In the observed cases, when the accretion rate is high ($\dot{m}\gtrsim 1$), the thin disc forms, and only a small fraction of black holes have jet \cite{Rafter2009AJ}. However, when the accretion rate is lower ($\dot{m}\lesssim1$), the ADAF forms, and the fraction of  black holes with jet increases dramatically \cite{Yang2020ApJ}. For most of the cases, our PBH has the accretion rate $\dot{m}$ well below $\sim1$, therefore it is reasonable to assume that a sufficient large fraction of them have jets.

Ref. \cite{Merloni2003} analyzed the selected samples including supermassive black holes and stellar-mass black holes with both X-ray and radio signals. They found a tight correlation between the radio luminosity $L_R$
($L_R=\nu_{5.0} L_{\nu_{5.0}}$, $\nu_{5.0}=5.0$ GHz), hard X-ray (2-10 keV) luminosity $L_X$ and black hole mass $M_\bullet$,
\begin{equation}
\log L_R=0.60\log L_X+0.78\log M_\bullet +7.33,
\label{eq:MHD03}
\end{equation}
where $L_R$ and $L_X$ are in the unit of ${\rm erg}\,{\rm s}^{-1}$, and  $M_\bullet$ is in units of ${\rm M}_\odot$. They pointed out that for models where the X-ray signal is  mainly from the standard thin disc and corona, or mainly from the jet, the predictions do not agree with the observed scaling relation. However,  if the radio signal is mainly from the jet and the X-ray signal is mainly from the ADAF accretion, such scaling relation is well interpreted \cite{Heinz2003MNRAS}.

We assume the PBH has radio from the jet and X-ray from the ADAF accretion. The radio luminosity and hard X-ray luminosity of the PBH also follow the scaling relation Eq. (\ref{eq:MHD03}). 
To derive the radio luminosity of the PBH from Eq. (\ref{eq:MHD03}), we use the following formula to calculate the hard X-ray luminosity 
\begin{equation}
\log (L_X/L_{\rm Edd})=\begin{cases}
2.3\log \dot{m}+1.1 ~~&\dot{m} \le 0.02\\
0.25\log \dot{m}-2.4~~&\dot{m} > 0.02.
\end{cases}
\end{equation}
The first line fits the outputs of theoretical calculations of the ADAF accretion \cite{Merloni2003}; the second line is derived from the bolometric correction of observed AGNs with accretion rate $\dot{m}\gtrsim0.02$, assuming a radiative efficiency 0.1 \cite{Lusso2012MNRAS}.

We assume the radio spectrum has a power-law form with a constant spectrum index
\begin{equation}
    L_{\nu}=\frac{L_R}{\nu_{5.0}}\left(\frac{\nu}{\nu_{5.0}}\right)^{-\alpha},
\end{equation}
where $\alpha=0.75$~\cite{blazars_0.75_1702.02571}.

\begin{figure}
\centering{
\subfigure{\includegraphics[width=0.45\textwidth]{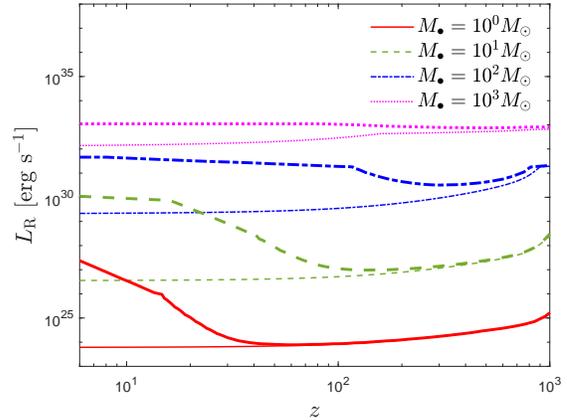}}
\caption{ The evolution of the radio luminosity at  5.0 GHz with redshift, for different PBH masses. Thin lines refer to the naked PBHs, while thick lines refer to PBHs surrounded by dark matter halos.
}
\label{fig:L_R_tot}
}
\end{figure}

In  Fig.~\ref{fig:L_R_tot}, we show the radio luminosity as a function of redshift for different PBH masses. For naked PBHs, $L_R$ is less evolved except at the highest redshifts. For PBHs inside dark matter halos, at high redshifts $L_R$ is close to the naked PBHs. However at lower redshifts it increases rapidly until the accretion rate reaching the Eddington limit. At $z\sim20$, when $M_{\bullet}=1-10^3\,{\rm M}_{\odot}$, $L_R\sim5 \times 10^{-15} -1 \times 10^{-9} L_{\rm Edd}$ for naked PBHs; and $L_R\sim8 \times 10^{-14} -9 \times 10^{-9} L_{\rm Edd}$ for PBHs inside dark matter halos.

\subsection{The cosmic radio background}\label{sec:Cosmic-Radio-Background}

To calculate the contribution of PBHs to the CRB, we need to know their mass distribution and number density. In this paper, for simplicity we assume that all PBHs have the same mass $M_{\bullet}$, then the comoving number density of PBHs is
\begin{equation}
    n_{\bullet}=\frac{f_{\rm PBH}\Omega_{\rm m}\rho_{\rm cr}}{M_{\bullet}},
\end{equation}
where $f_{\rm PBH}$ is the ratio of total PBH  mass density to total matter density in the Universe, and $\Omega_{\rm m}$ is the fractional matter density.
 
The CRB from PBHs is then
\begin{equation}
\begin{aligned}
    J_{\rm CRB}(\nu,z) &=\frac{(1+z)^3}{4\pi}\int_z^{\infty}\epsilon(\nu^\prime,z^\prime)\frac{c}{H(z^\prime)(1+z^\prime)}{\rm d}z^\prime \\
    &=\frac{(1+z)^3}{4\pi}\int_z^{\infty}\frac{ f_{\rm DC}  n_{\bullet}L_{\nu'}(M_{\bullet}) c}{H(z^\prime)(1+z^\prime)}{\rm d}z^\prime,
\end{aligned}
\end{equation}
where 
\begin{equation}
    \nu^\prime=\nu \left(\frac{1+z^\prime}{1+z}\right),
\end{equation}
is the frequency in the emitter's rest frame. $f_{\rm DC}\le 1$ is the duty-cycle for radio emission, which refers to the fraction of PBHs that is active in the radio band. Since the radio emission for these PBHs is produced by the jet, it would be equal to the ratio of the jet lifetime compared to the Hubble timescale.

\begin{figure*}
    \centering
    {
    \subfigure{\includegraphics[width=0.45\textwidth]{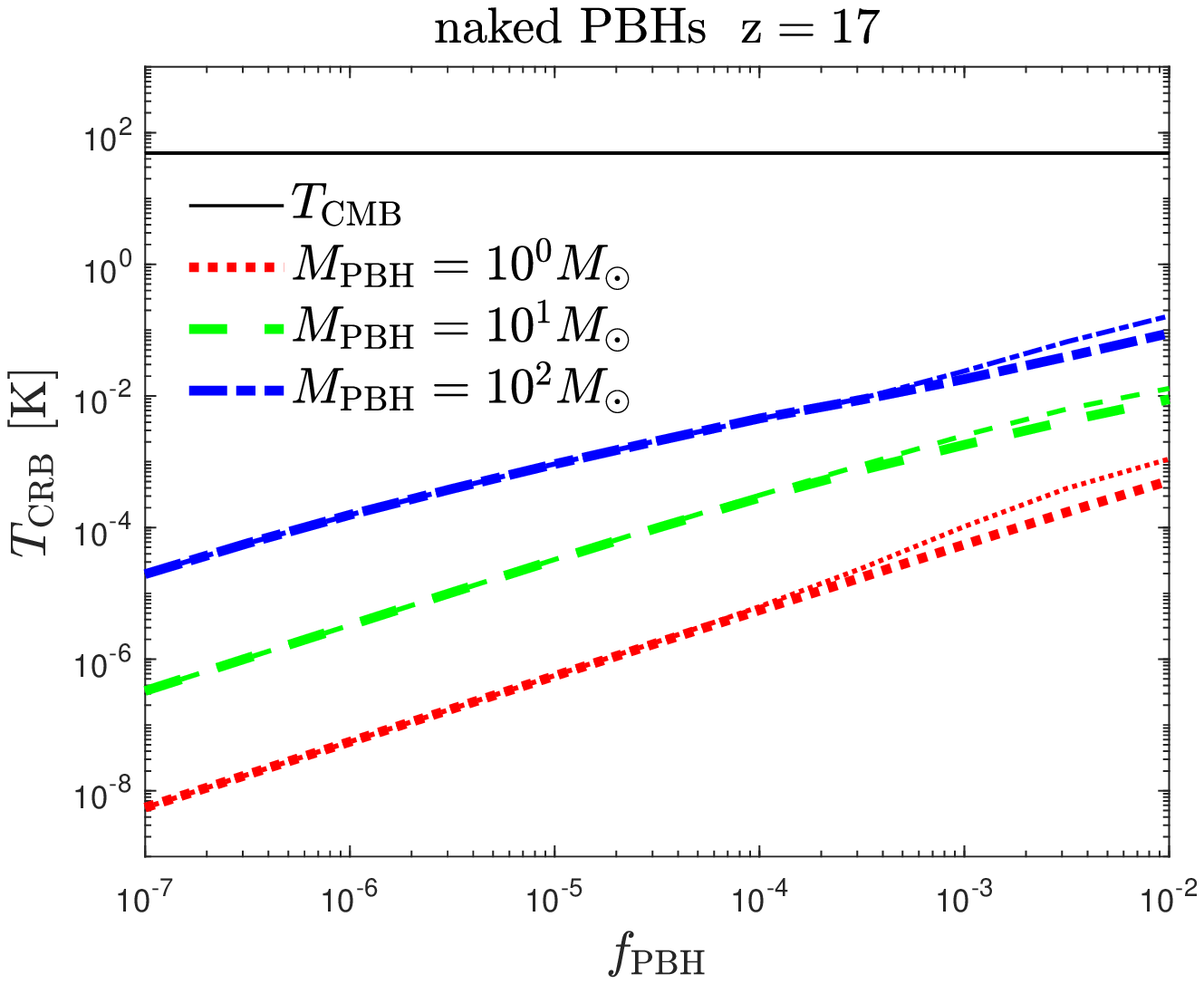}}
    \subfigure{\includegraphics[width=0.45\textwidth]{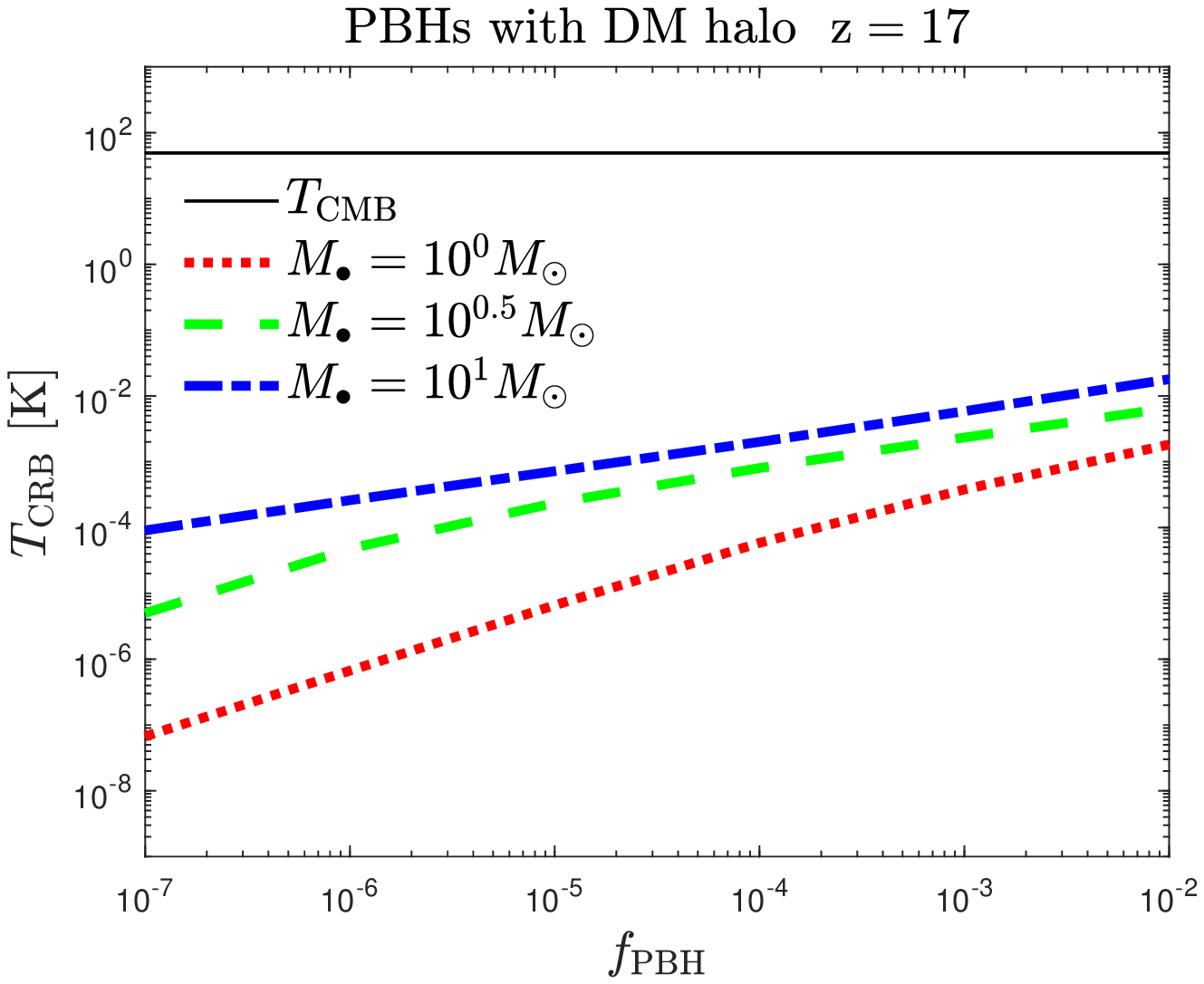}} 
    \caption{The contribution to the CRB at $z = 17$  for various PBH masses, as a function of $f_{\rm PBH}$. {\it Left}: Naked PBHs; {\it Right}: PBHs inside dark matter halos. 
    Since the X-ray heating can reduce the accretion rate, $T_{\rm CRB}$ is not necessarily proportional to $f_{\rm PBH}$. In the left panel, the  thick (thin) lines refer to the model without (with) PBH binaries.
    }
    
    \label{fig:T_CRB_new1}
    }
\end{figure*}

In Fig.~\ref{fig:T_CRB_new1} we show the contribution to the 1.4 GHz CRB at $z\sim17$ from PBHs as a function of $f_{\rm PBH}$, for naked PBHs and for PBHs inside dark matter halos respectively. However, even if we set $f_{\rm DC}=1$, this contribution is still negligible compared with the CMB. In principle, if $M_{\bullet}$ is higher and/or $f_{\rm PBH}$ is higher, one may obtain higher $T_{\rm CRB}$, but the X-ray heating (see Sec.~\ref{sec:X-ray-heating}) can reduce the accretion rate, so $T_{\rm CRB}$ increases more gently for higher $M_{\bullet}$ at high $f_{\rm PBH}$ region. More importantly, the X-ray heating works more efficiently in reducing the 21-cm absorption than the extra CRB in boosting the 21-cm absorption.
We have checked that using a higher PBH mass and a higher abundance than in Fig.~\ref{fig:T_CRB_new1} will not help boosting the 21-cm absorption signal.

    \begin{figure*}
    \centering
    {
    \subfigure{\includegraphics[width=0.45\textwidth]{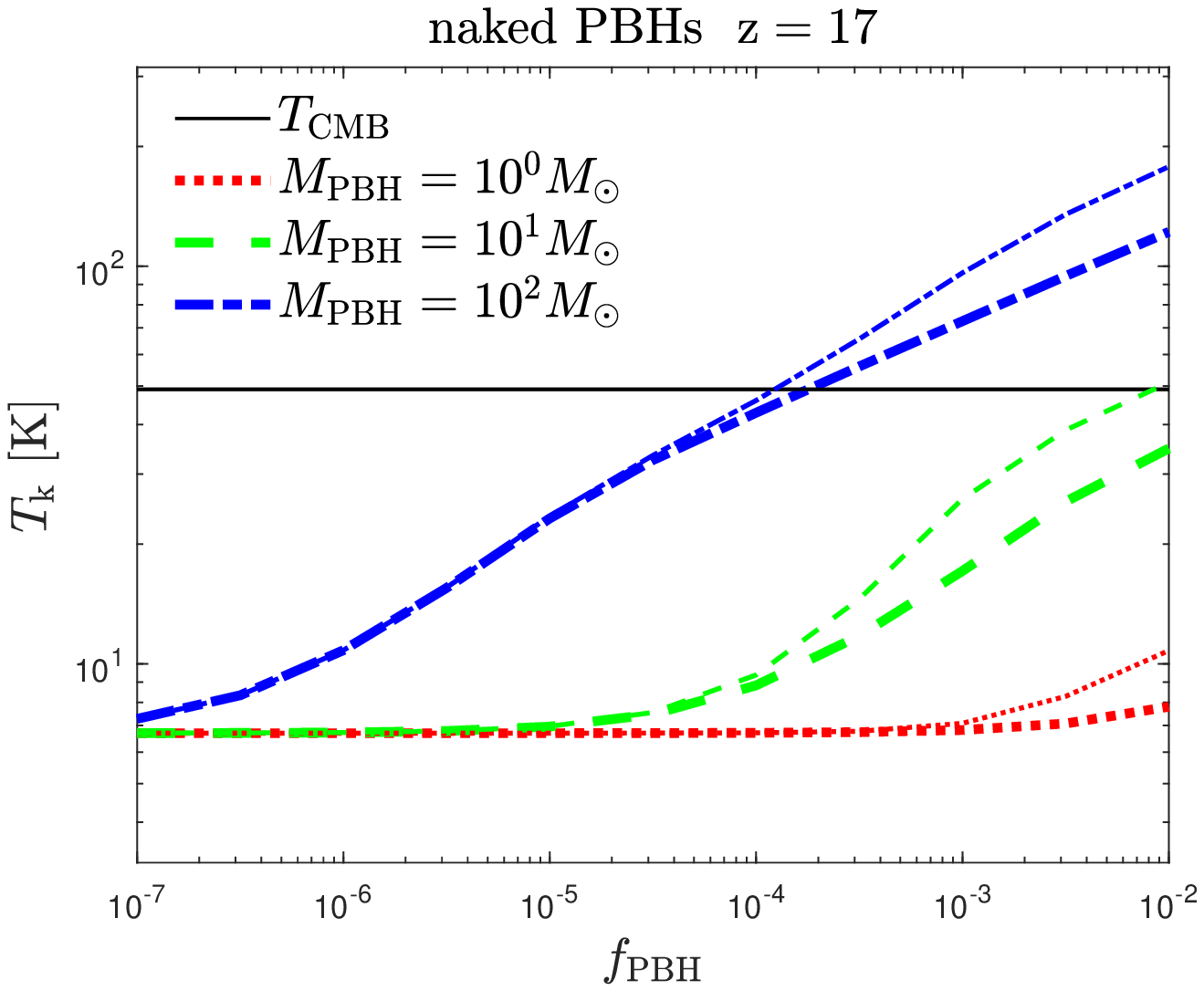}}
    \subfigure{\includegraphics[width=0.45\textwidth]{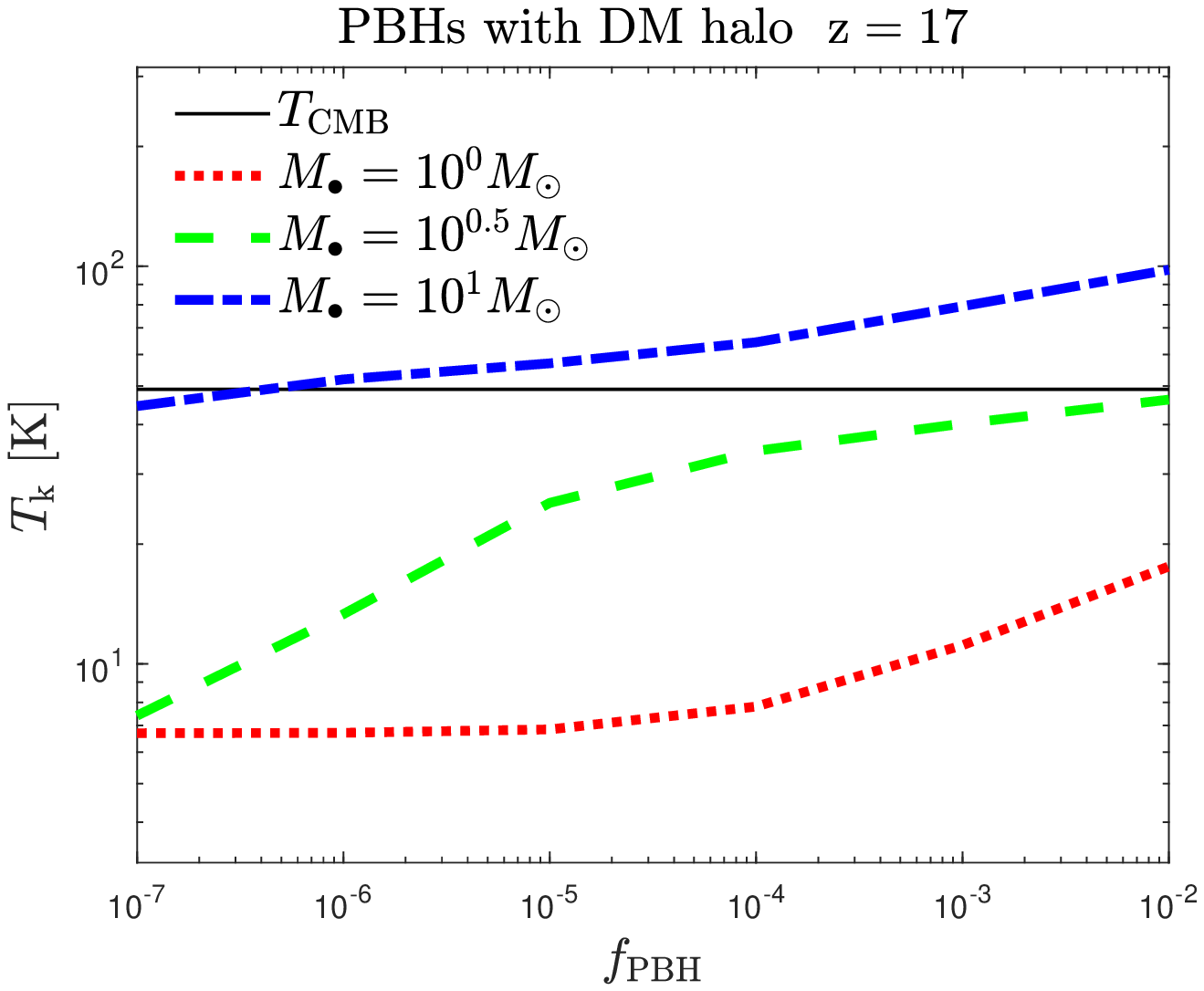}} 
    \caption{The kinetic temperature of the IGM at $z= 17$ heated by PBHs with various masses, as a function of $f_{\rm PBH}$. {\it Left}: Naked PBHs; {\it Right}: PBHs inside dark matter halos. In the left panel, the  thick (thin) lines refer to the model without (with) PBH binaries.
    }
    
    \label{fig:T_k_new1}
    }
\end{figure*}

\subsection{The X-ray heating and ionization of the IGM}\label{sec:X-ray-heating}

PBHs emit not only radio photons but also X-ray photons. They can heat and ionize the IGM, which would reduce the 21-cm absorption signal.

The {\it physical} luminosity density of X-ray emission that interacts with the IGM is
\begin{equation}
l_X= n_{\bullet}(1+z)^3 \int_{E_{\rm min}}^\infty  L_X(E)[1-e^{-\tau(E)}]  {\rm d}E, 
\end{equation}
where the optical depth $\tau(\nu)=[\sigma_{\rm H}(E)n_{\rm H}(1-x_{\rm HII}) +\sigma_{\rm He}(E) n_{\rm He}(1-x_{\rm HeII})](c/H)$, and $\sigma_{\rm H}$ and $\sigma_{\rm He}$ are photon-ionization cross-sections for hydrogen and helium respectively~\cite{CRASH2003}. 
Here we assume that the X-ray spectrum follows the form 
\begin{equation}
L_X(E)=L_{0} \left( \frac{E}{E_{\rm min}} \right)^{-0.5} \exp(-E/E_1)   
\end{equation}
and adopt $E_{\rm min}=2$ keV and $E_{1}=100$ keV,
with the normalization    
\begin{equation}
L_X=\int_{\rm 2\,keV}^{\rm 10\,keV}L_X(E) {\rm d}E.    
\end{equation}

Ignoring the contribution from UV photons, the time evolution of hydrogen ionized fraction $x_{\rm HII}$, helium ionized fraction $x_{\rm HeII}$,   and temperature of the IGM $T_{\rm k}$ follows
\begin{eqnarray}
\frac{{\rm d} x_{\rm HII}}{{\rm d}t}
&=& I_{\rm H} +\chi_{\rm ion,H} \frac{l_X}{n_{\rm H}E_{\rm H} } \nonumber  \\
\frac{{\rm d} x_{\rm HeII}}{{\rm d}t}
&=& I_{\rm He} +\chi_{\rm ion,He} \frac{l_X}{n_{\rm He}E_{\rm He} } \nonumber  \\ 
\frac{{\rm d} T_{\rm k} }{{\rm d}t}&=& H_T +\frac{2}{3k_{\rm B} n}\chi_{\rm heat}l_X,
\end{eqnarray} 
where $n_{\rm H}=(1-Y_{\rm He})\Omega_{\rm b}\rho_{\rm cr}(1+z)^3/m_{\rm H}$ and $n_{\rm He}=Y_{\rm He}\Omega_{\rm b} \rho_{\rm cr}(1+z)^3/m_{\rm He}$ are {\it physical} number density of hydrogen and helium elements respectively, and $Y_{\rm He}=0.24$;
$x_{\rm e}= x_{\rm HII}+n_{\rm He}/n_{\rm H}  x_{\rm HeII} $; 
the total number density of particles $n=n_{\rm H}(1+x_{\rm e})+n_{\rm He}$; $E_{\rm H}=13.6$ eV and $E_{\rm He}=24.6$ eV are ionization energy for hydrogen and helium; $k_{\rm B}$ is the Boltzmann constant; $\chi_{\rm ion,H}$,  $\chi_{\rm ion,He}$  and  $\chi_{\rm heat}$ are fractions of X-ray energy deposited into hydrogen ionization, helium ionization and heating
\cite{VF08};
the net ionization ($I_{\rm H}$ and $I_{\rm He}$) and heating ($H_{T}$) rates in the absence of PBH are:
\begin{eqnarray}
I_{\rm H}&=&  [\gamma_{\rm H}(T_{\rm k})(1-x_{\rm HII}) -\alpha_{\rm H}(T_{\rm k})x_{\rm HII}] x_{\rm e} n_{\rm H} \nonumber \\
I_{\rm He}&=&  [\gamma_{\rm He}(T_{\rm k})(1-x_{\rm HeII}) -\alpha_{\rm He}(T_{\rm k})x_{\rm HeII}] x_{\rm e} n_{\rm H} \nonumber \\
H_T&=& -2H(z)T_{\rm k} -\frac{1}{n} \left[n_{\rm H} \frac{{\rm d} x_{\rm HII}}{{\rm d}t}+n_{\rm He}\frac{{\rm d} x_{\rm HeII}}{{\rm d}t}\right]T_{\rm k} \nonumber \\
&&-\frac{2}{3k_{\rm B} n}\Lambda(T_{\rm k}), \nonumber \\
\end{eqnarray}
where $\gamma_{\rm H, He}$ and $\alpha_{\rm H, He}$ are rates of collision ionization and recombination, for hydrogen and helium respectively~\cite{CRASH2003}; 
$\Lambda(T)$ are the {\it physical volume} cooling rate  (including the  collisional  ionization cooling, collisional excitation, Bremsstrahlung cooling and Compton heating/cooling)~\cite{CRASH2003}. Since $\dot{m}$ depends on $T_{\rm k}$ and $x_{\rm e}$, we solve the thermal evolution of the IGM and $\dot{m}$ self-consistently.  

In Fig.~\ref{fig:T_k_new1}, we show the kinetic temperature of the IGM at $z\sim17$ as a function of $f_{\rm PBH}$, for various PBH masses same as in Fig.~\ref{fig:T_CRB_new1}, for naked PBHs and PBHs inside dark matter halos respectively.
In both cases,  it is easier to heat the IGM temperature above the CMB than to generate the CRB larger than the CMB.

\begin{figure*}
    \centering   {\subfigure{\includegraphics[width=0.45\textwidth]{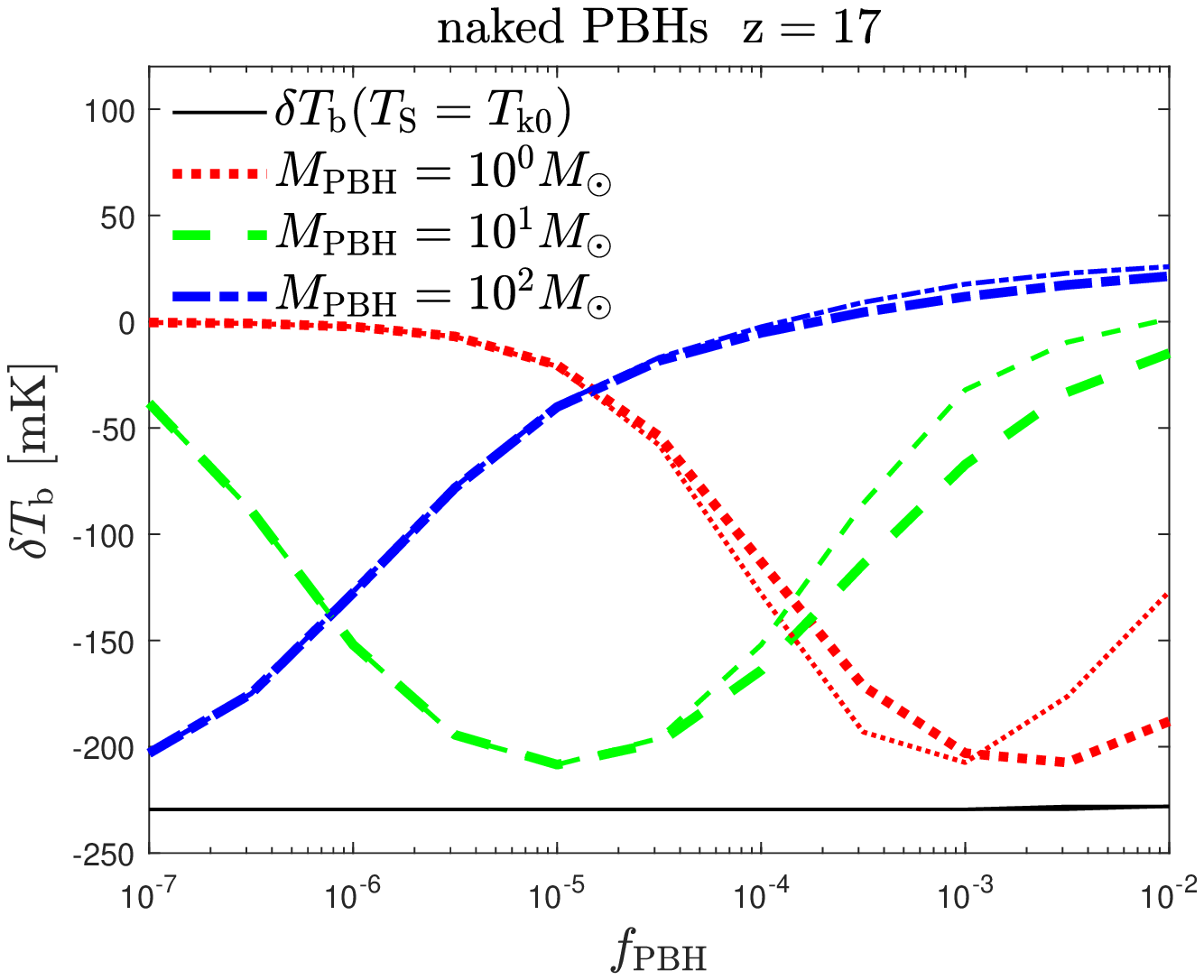}}
    \subfigure{\includegraphics[width=0.45\textwidth]{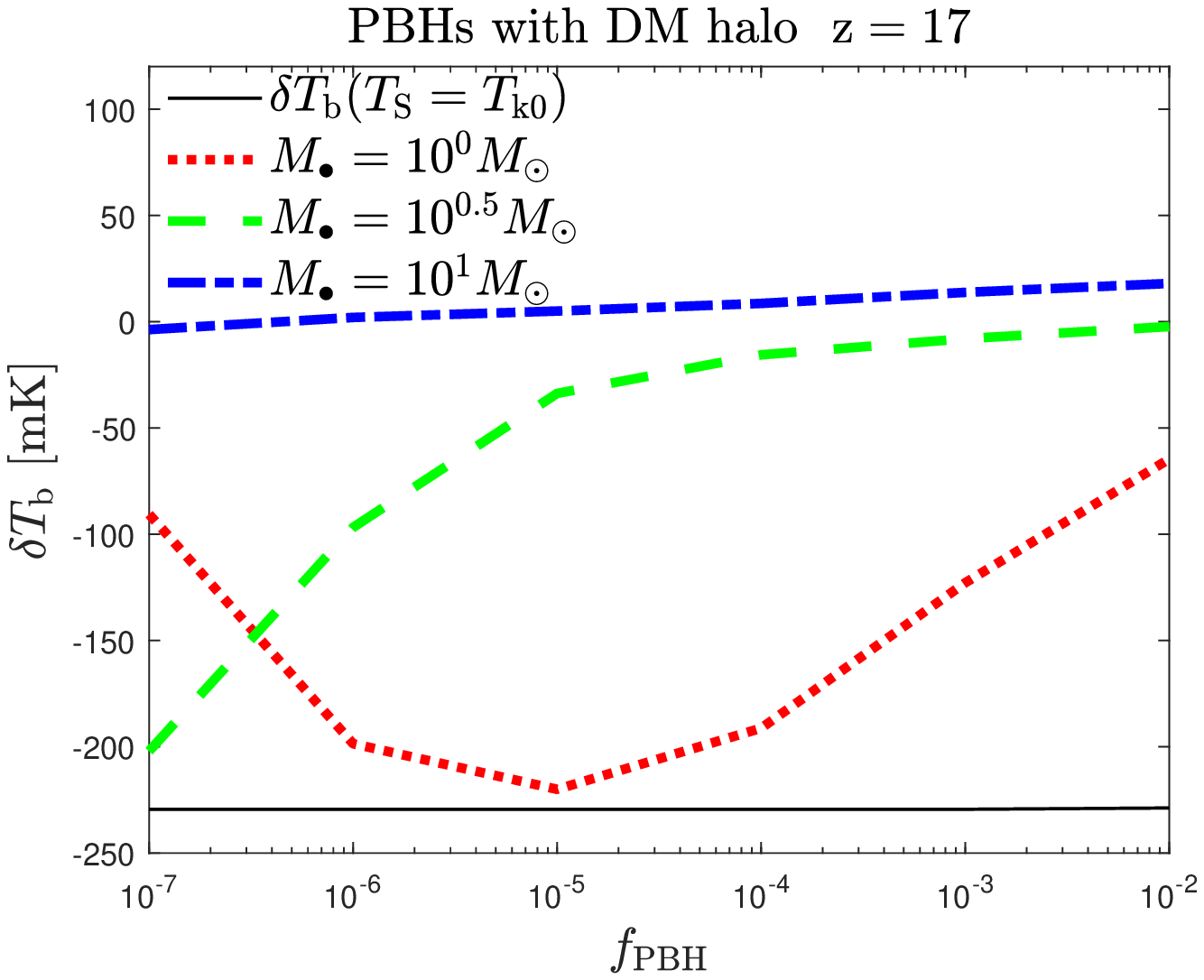}} 
    \caption{
    The 21-cm brightness temperature at $z \sim 17$ under the influence of PBHs with different masses, as a function of PBH abundance. {\it Left}: Naked PBHs; {\it Right}: PBHs surrounded by dark matter halos. 
    The black solid line represents  the maximum 21-cm absorption, i.e. $-229$ mK, at $z\sim17$ in the standard cosmology.
    However, please note that the inhomogeneity of the IGM and shock heating effect can reduce this maximum absorption by $\sim15\%$~\cite{Xu2018ApJ,EDGES_stronger_2102.12865}. In the left panel, the  thick (thin) lines refer to the model without (with) PBH binaries.
    }
    \label{fig:T_21_new1}
    }
\end{figure*}

\subsection{The global spectrum of 21-cm signal}

The brightness temperature of 21-cm global spectrum is~\cite{21cm_review2_2012RPPh...75h6901P}:
\begin{eqnarray}
\delta T_{\rm b} &=& 27 x_{\rm HI} \left(\frac{\Omega_{\rm b} h^2}{0.023}\right)\left( \frac{0.15}{\Omega_{\rm m} h^2} \frac{1+z}{10} \right)^{1/2} \nonumber \\
& \times & \left( \frac{T_{\rm S}-T_{\rm R}}{T_{\rm S}}\right)~[\rm mK],
\end{eqnarray}    
where the temperature of the radiation field $T_{\rm R}=T_{\rm CMB}+T_{\rm CRB}$.
The spin temperature is
\begin{equation}
T_{\rm S}^{-1}=\frac{T_{\rm R}^{-1} +x_\alpha T_{\rm k}^{-1} +x_{\rm c}T_{\rm k}^{-1}}{1+x_\alpha+x_{\rm c}},    
\end{equation}
where $x_\alpha$ and $x_{\rm c}$ are coupling coefficients due to Ly$\alpha$ scattering and collision respectively. Calculations of them can be found in Ref.~\cite{21cmFAST2011MNRAS} and references therein. 

The PBH does not only generate radio background and X-ray which heats up the gas, but also contribute to the  
Ly$\alpha$ background. A fraction $\chi_\alpha$\cite{VF08} of the X-ray energy is deposited into Ly$\alpha$ photons, so the specific intensity of photon number is given by 
\begin{equation}
J_\alpha(z)=\frac{1}{h_{\rm P}\nu_\alpha}\frac{1}{4\pi} \frac{ \chi_\alpha l_X}{\nu_\alpha } \frac{c}{H(z)}.
\end{equation}

In Fig.~\ref{fig:T_21_new1}, we show the 21-cm brightness temperature at $z=17$ as  a function of  $f_{\rm PBH}$, for various PBH masses. To highlight the role played by the PBH, we ignore the radiation from any normal astrophysical objects, e.g. Pop III stars, first galaxies.
It is interesting to note that, although the CRB from the PBH is unimportant in this scenario, for the 21-cm absorption there is an optimal $f_{\rm PBH}$ where the absorption reaches the maximum. If $f_{\rm PBH}$ is lower, it produces less Ly$\alpha$ background, which is not strong enough to couple $T_{\rm S}$ with $T_{\rm k}$, therefore the 21-cm absorption is weaker. If $f_{\rm PBH}$ is higher however, the IGM is heated significantly, which still results in weaker 21-cm absorption. 
Even in the absence normal astrophysical objects, the PBHs can provide the necessary  Ly$\alpha$ photons to make the 21-cm signal be close to the maximum absorption in the standard cosmology. This is not easy if only the  Pop III stars are the first ionization sources~\cite{PopIII1,21cm_2006MNRAS.371..867F,PopIII3}.

\subsection{The PBH binaries}
In addition to the isolated PBHs, there are also PBH binaries. Ref. \cite{DeLuca2020JCAP} found that: for two PBHs with equal mass, if they form a binary system, their total accretion rate is the Bondi accretion rate of a single black hole with mass equal to the sum of their mass. Since Bondi accretion rate is generally proportional to  $\sim M^2_\bullet$, if the indirect dependence of $\lambda$  on  $M_\bullet$ is ignored,
the accretion rate of each PBH is then boosted by a factor  $\sim$2, because in the binary system, the ambient medium density of each PBH is boosted. 
The boost factor of radiation (X-ray, radio) is even larger than 2 since the bolometric luminosity is proportional to $\dot{m}^2$.
Therefore, the PBH binaries can help to increase the influence on the 21 cm signal.
Ref. \cite{DeLuca2020JCAP} also pointed out that,  the binary system will finally reach a stationary state where the two members have equal mass, even if at the beginning their mass is different, because the accretion evolution tends to balance each other. 

On the other hand, the fraction of PBHs that are in binary systems is \cite{Raidal2017JCAP}
\begin{equation}
f_{\rm binary}\sim 1-e^{-\delta_{\rm dc}f_{\rm PBH}},
\end{equation}
where $\delta_{\rm dc}$ describes the clustering of PBHs. For uniform distribution, $\delta_{\rm dc}=1$. In such case, $f_{\rm binary}$ is small when $f_{\rm PBH} \lesssim 0.1$, and the contribution to X-ray heating and CRB from PBH binaries is small. However, if for some reason the PBHs can be highly clustered, say $\delta_{\rm dc}\gtrsim10$, then the PBH binaries may have a significant influence on the 21 cm signal, depending on $f_{\rm PBH}$. In the left panels of Fig. \ref{fig:T_CRB_new1}, \ref{fig:T_k_new1} \& \ref{fig:T_21_new1} we show the $T_{\rm CRB}$, $T_{\rm k}$ and $\delta T_{\rm b}$ of 21 cm signal if the contribution from PBH binaries is also taken into account by thin lines, assuming $\delta_{\rm dc}=1000$. 
We see that when $f_{\rm PBH}\gtrsim 10^{-4}-10^{-3}$, the influence of PBH binaries becomes more obvious.
The gravitational wave observations put constraints on the joint parameter $\delta_{\rm dc}^{16/37} f_{\rm PBH}^{53/37}$. For $\delta_{\rm dc}=1000$, $f_{\rm PBH}\sim 10^{-4}-10^{-3}$ is still within the limit set by LIGO observations \cite{Raidal2017JCAP}. 

If the PBH binaries are located inside dark matter halos, absolutely they will also boost the influence on the 21 cm absorption signal. We do not show the results in the right panels because we suppose the accretion rate, in this case, can be much more complicated than the binaries of naked PBHs.

\section{The CRB and 21-cm signal from the jet model}\label{sec:jet}

So far we have calculated the radio and X-ray radiation from PBHs with the empirical relation Eq. (\ref{eq:MHD03}).  
However, it is also possible that the PBHs do not necessarily follow this scaling relation, since they were born in a different environment. 
Moreover, it is pointed out that in Ref.  \cite{Merloni2003} different sample objects may have different X-ray origins. For the low accretion rate objects, 
both the radio and X-ray signals can arise predominantly from the jet \cite{Mezcua2018MNRAS}. In this section, we tentatively calculate the radio emission using a jet model that can produce a stronger radio signal, then re-evaluate the CRB and 21-cm signal at Cosmic Dawn under reasonable assumptions about the model parameters.

\subsection{The jet model }

When accretion is weak, the radiative efficiency of the accretion is low, but the relativistic electrons in the jet can produce strong synchrotron  radiation~\cite{Yuan2005,Jet3_1711_01292}.
In this case we can derive the radio luminosity from a jet model where all jet emission is considered as coming from a blob of electrons with uniform properties \citep{Ghisellini2010}.  

The total synchrotron power of the blob is \citep{Ghisellini2013}
\begin{equation}
P_{\rm syn}=\int_{\gamma_{\rm min}}^{\gamma_{\rm max}} \frac{4}{3}\sigma_{\rm T}c \gamma^2 U_B \frac{d N_{\rm e}(\gamma)}{{\rm d}\gamma}{\rm d}\gamma,
\end{equation}
where ${\rm d}N_{\rm e}/{\rm d}\gamma$ is the energy  distribution of relativistic electrons in the blob; $U_B=B^2/(8\pi)$ is the energy density of magnetic field with strength $B$; $\gamma_{\rm min}$  and $\gamma_{\rm max}$ are the minimum and maximum Lorentz factors respectively. 

On the other hand, since the jet blob is far from the central black hole, in redshift ranges of our interests, the seed photons for inverse Compton scattering (ICS) are mainly the synchrotron photons (the synchrotron self-Compton, SSC) and the CMB photons. The total ICS power is
\begin{align}
P_{\rm ICS}= \int_{\gamma_{\rm min}}^{\gamma_{\rm max}}\frac{4}{3}\sigma_{\rm T}   c \gamma^2 U_{\rm R} \frac{{\rm d}N_{\rm e}(\gamma)}{{\rm d}\gamma} {\rm d}\gamma,
\end{align}
where $U_{\rm R}=\rho_{\gamma}+U_{\rm syn}$ is the energy density of radiation field, consisting of CMB energy density ($\rho_{\gamma}$) and synchrotron radiation field. The latter is 
\begin{eqnarray}
U_{\rm syn} \sim \frac{P_{\rm syn}R_{\rm blob}/c}{4\pi R_{\rm blob}^3/3},    
\end{eqnarray}
where $R_{\rm blob}$ is the physical size of the blob.

\begin{figure}
	\centering
	\includegraphics[width=0.45\textwidth]{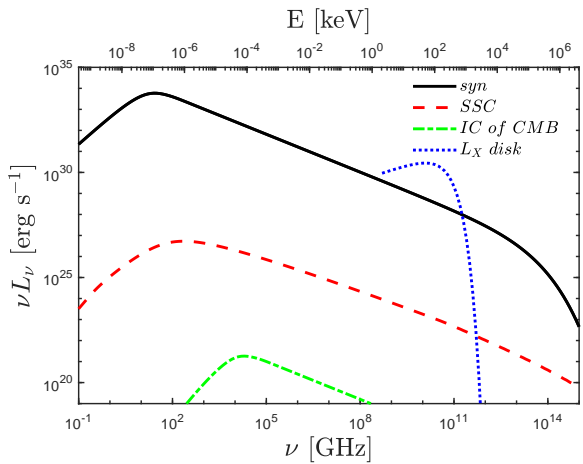}
	\caption{The different components of the SED of a PBH with $M_{\bullet}=10\,{\rm M}_{\odot}$ at $z=17$.}
	\label{fig:SED}
\end{figure}

It is generally assumed that the injected relativistic electrons follows a power-law distribution with an exponential cutoff (e.g.~\cite{Pepe2015AA}), 
\begin{equation}
Q(\gamma)=Q_0\gamma^{-s}\exp\left(-\frac{\gamma}{\gamma_{\rm max}}\right),    
\end{equation}
where the normalization $Q_0$ is set by the relation
\begin{eqnarray}
P_{\rm e}=\int  m_{\rm e} c^2 \gamma Q(\gamma) {\rm d}\gamma.
\end{eqnarray}
$P_{\rm e}=\eta_{\rm e} \eta_{\rm jet}\dot{m}L_{\rm Edd}$ is the injected power of the relativistic electrons, where $\eta_{\rm jet}$ is the jet efficiency and $\eta_{\rm e}$ is the fraction of jet power that is finally converted into the relativistic electrons.

With the energy loss due to synchrotron and ICS, the steady state electron distribution is~\cite{Ghisellini2009MNRAS,Gardner2014MNRAS,Gardner2018MNRAS}
\begin{eqnarray}
 \frac{{\rm d}N_{\rm e}(\gamma)}{{\rm d}\gamma}=
 \begin{cases}
 KQ(\gamma)&\gamma < \gamma_{\rm cool} \\
 -\frac{1}{\dot{\gamma}} \int_{\gamma}^{\gamma_{\rm max}}Q(\gamma'){\rm d}\gamma'& \gamma > \gamma_{\rm cool},
 \end{cases}, \label{eq:dNe_dgamma}
\end{eqnarray}
where
\begin{eqnarray}
\dot{\gamma}=-\frac{1}{m_{\rm e} c^2} \frac{4}{3}\sigma_{\rm T} c\gamma^2(U_{\rm R}+U_B).
\end{eqnarray}
$\gamma_{\rm cool}$ is the point where the cooling time equals the crossing timescale of the blob, with the cooling time given by
\begin{equation}
t_{\rm cool}=\frac{ \gamma m_{\rm e} c^2  }{\frac{4}{3}\sigma_{\rm T} c \gamma^2(U_{\rm R}+U_B)}
\end{equation}
and  $t_{\rm cross}=R_{\rm blob}/c$. $K$ factor in Eq.~(\ref{eq:dNe_dgamma}) is
\begin{eqnarray}
K=-\frac{1}{\dot{\gamma} Q(\gamma)} \int_{\gamma}^{\gamma_{\rm max}}Q(\gamma'){\rm d}\gamma'\Big|_{\gamma=\gamma_{\rm cool}}.
\end{eqnarray}

\begin{figure*}
\centering{
\subfigure{\includegraphics[width=0.3\textwidth]{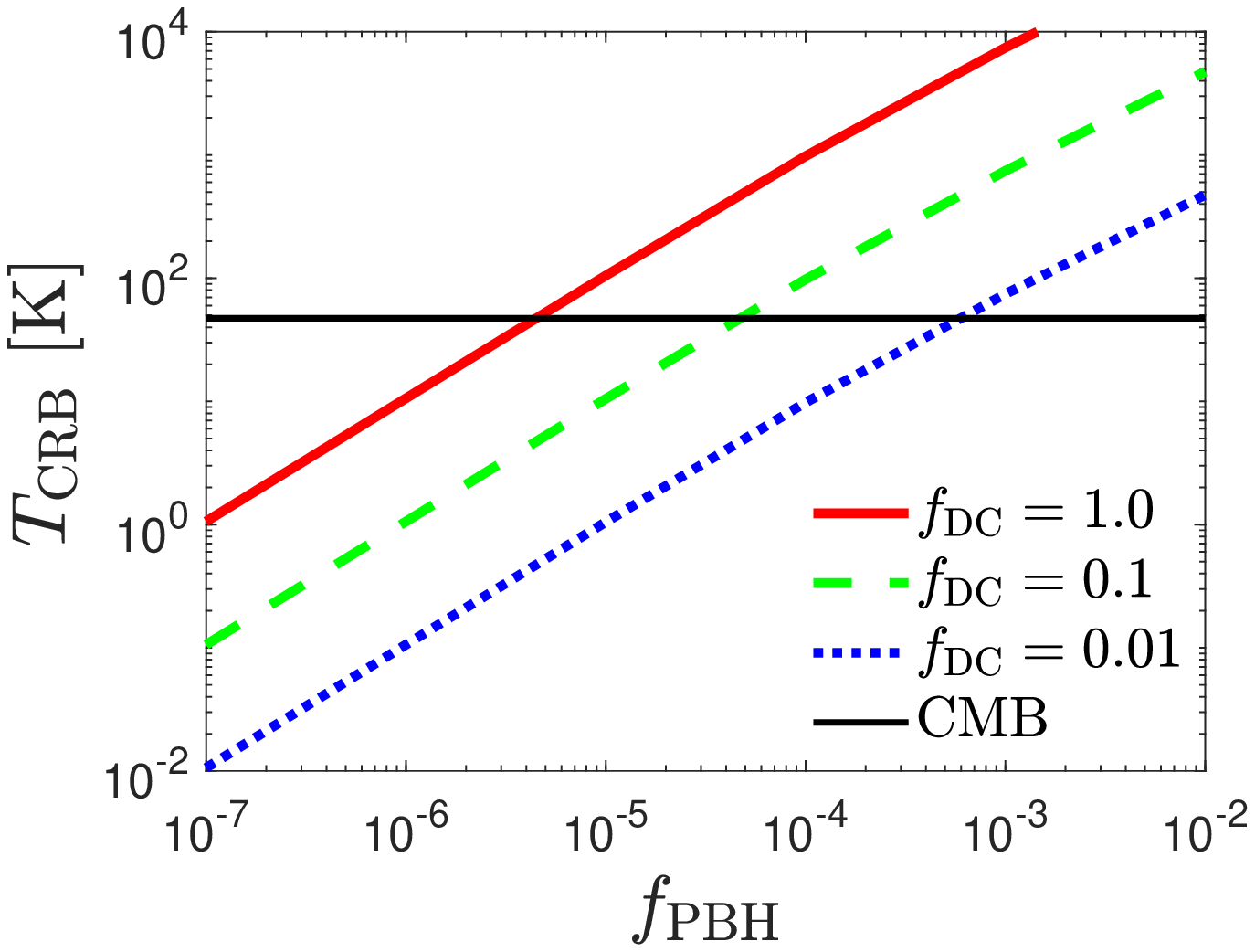}}
\subfigure{\includegraphics[width=0.3\textwidth]{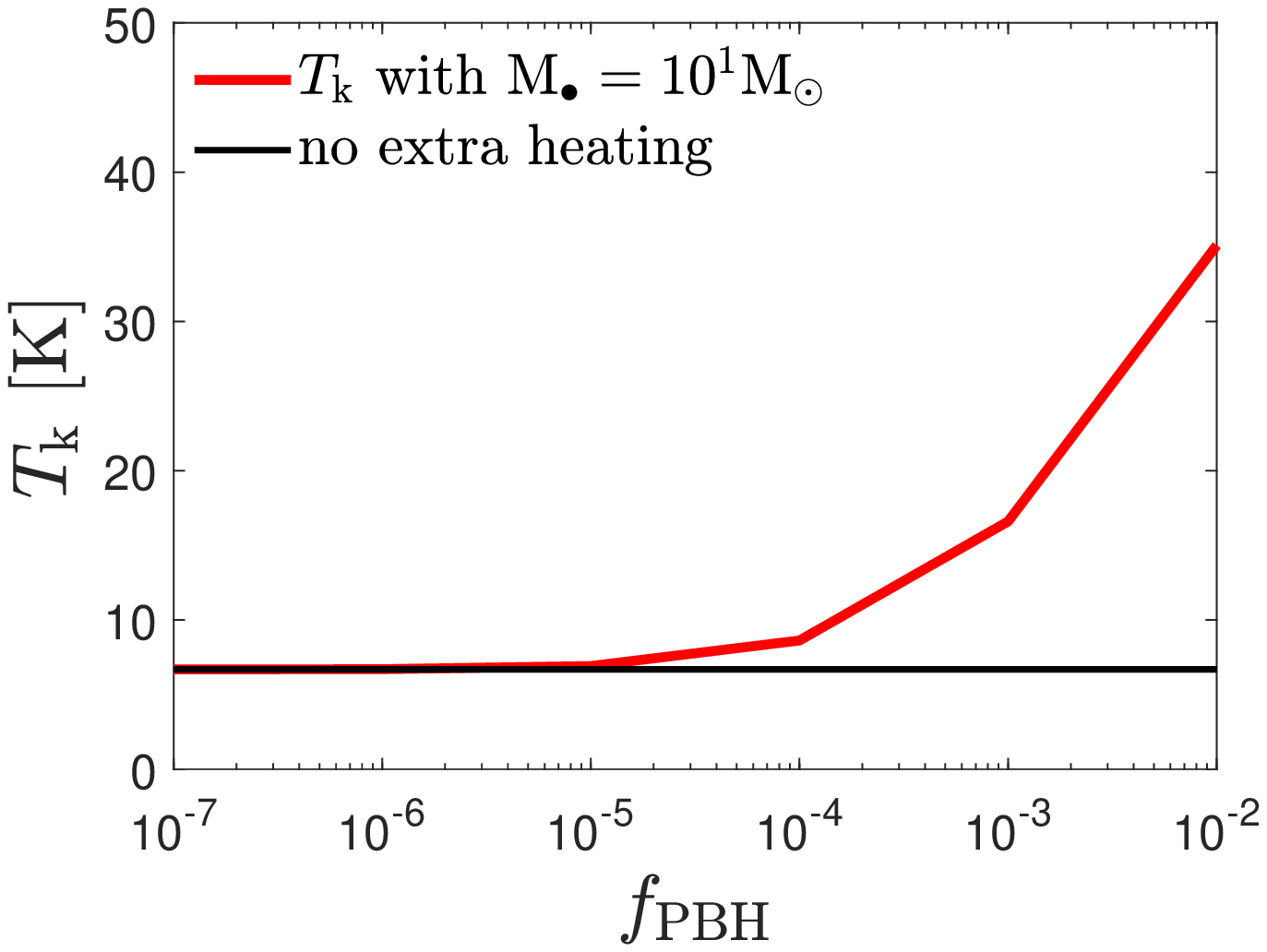}}
\subfigure{\includegraphics[width=0.3\textwidth]{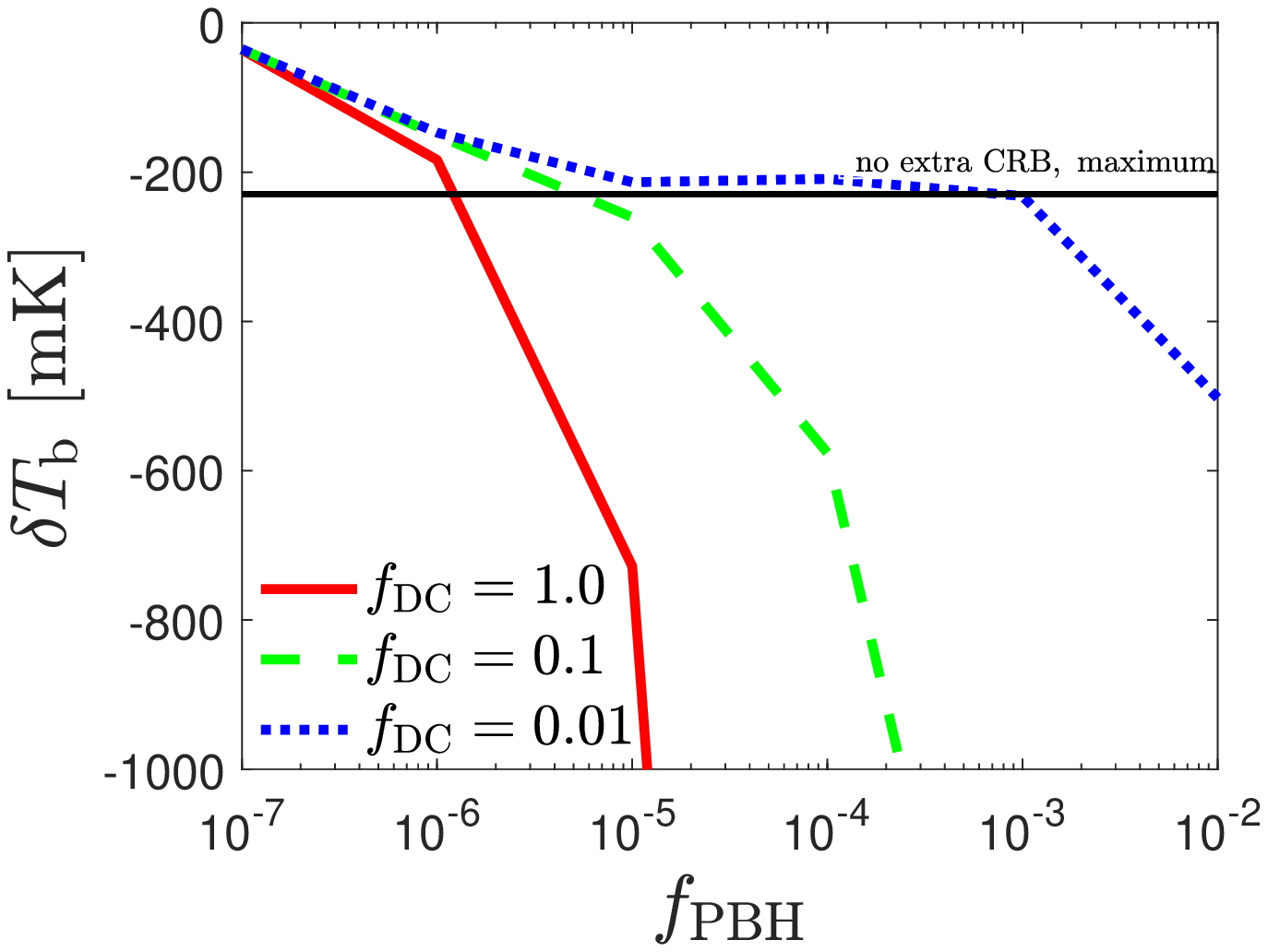}}
\caption{The CRB temperature from PBHs (left), the IGM temperature (middle); and the 21-cm signal (right), at $z=17$ as a function of PBH abundance. We show the results for three different $f_{\rm DC}$. We assume all PBHs have the same mass of $10\,{\rm M}_{\odot}$,  and fiducial jet parameters are adopted. For comparison, in the CRB panel we plot the CMB temperature; in the kinetic temperature panel we plot the IGM temperature in the absence of extra heating; in the 21-cm signal panel we plot the maximal absorption signal in the absence of extra CRB and heating. All quantities are calculated for naked PBHs. } 
\label{fig:temperatures}
}
\end{figure*}

In practice, $\gamma_{\rm cool}$ is found by numerical iterations.
Finally the synchrotron luminosity is
\begin{equation}
L_{\rm syn}(\nu)=\int_{\gamma_{\rm min}}^{\gamma_{\rm max}} 
l_{\rm syn}(\nu,\gamma,B) \frac{{\rm d}N_{\rm e}(\gamma)}{{\rm d}\gamma} {\rm d}\gamma,    
\end{equation}
where $l_{\rm syn}$ is the synchrotron emission of a single electron~\cite{Ghisellini2013}.
We then obtain the SSC luminosity from $L_{\rm syn}(\nu)$ and ${\rm d}N_{\rm e}(\gamma)/{\rm d}\gamma$, and  the luminosity of the ICS between electrons and CMB photons~\cite{Lefa2012PhDT} from ${\rm d}N_{\rm e}(\gamma)/{\rm d}\gamma$.

So the jet model has following parameters $\{ R_{\rm bolb}, B,\eta_{\rm jet}, \gamma_{\rm min},\gamma_{\rm max}, s\}$ that are directly relevant to the radio signal. We assume that the PBH jet is analogous to the jet of typical stellar-mass black hole Cygnus X-1~\cite{Pepe2015AA,Gandhi2017NatAs}, 
in spite of its lower accretion rate.
We set the jet parameters according to the following considerations: 

\begin{itemize}

\item 
For Cygnus X-1, according to Ref. \cite{Zdziarski2012MNRAS,Heinz2006ApJ}, the bulk of the low frequency radio emission is generated at the height $\mathcal{O}(1)\times10^7~R_{\rm S}$ along the jet axis, where  $R_{\rm S}= 2GM_{\bullet}/c^{2}$ is the Schwarzschild radius. We assume  half of the opening angle of the blob is 0.1 radian, therefore take $R_{\rm blob}=10^6~R_{\rm S}$. 

\item According to Ref. \cite{Pepe2015A&A}, at the jet base which is $1.1\times 10^8$ cm above the back hole, $B=5\times 10^7$ G. $B$ can be inversely proportional to the height along jet axis, therefore at the end of acceleration region which is $\sim\mathcal{O}(1)\times 10^{12}$ cm above the black hole, $B$ is serval thousand G.
According to Ref. \cite{Zdziarski2014MNRAS}, at the onset of the energy dissipation $B\sim10^4-10^6$ G. If most of the radio emission is produced at the height $10^3$  times the onset, and $B$ is again inversely proportional to the height, then $B$ can be several tens to several thousand.  We take $B=2000$ G as the fiducial value. We  will also show the results for much lower $B$.

\item The observed total jet power can be comparable or even larger than the total accretion power $\dot{m}L_{\rm Edd}$ \cite{Ghisellini2014Natur}, because the jet power is not directly from the accretion \cite{Ghisellini2019RLSFN}, but supplied by the black hole spin. This is supported by the GRMHD simulations which found that the steady-state jet efficiency $\eta_{\rm jet}$ can be $\sim$ 140\% \cite{Tchekhovskoy2011MNRAS} for the ADAF-like accretion mode. 
The maximum jet efficiency can reach $\sim 300$\% \cite{McKinney2012MNRAS}.
Here we assume that a large fraction of the jet power is finally converted into the kinetic energy  of the relativistic electrons, and take $\eta_e\eta_{\rm jet}=0.5$ as the optimistic fiducial value.
We will also show the results for a much smaller $\eta_e\eta_{\rm jet}$.

\item Regarding the parameters of energy distribution of relativistic electrons, we take $\gamma_{\rm min}=1$ and $\gamma_{\rm max}\sim 10^6$ as in Ref. \cite{Zhang2014ApJ}. However we take a steeper slope $s=3.2$ as fiducial value, so that the low frequency signal is larger in the jet radiation. In some blazars the distribution of relativistic electrons with $\gamma \gtrsim 100$ can have such steep slope \cite{Ghisellini2010}. 
We will also show the results for a shallower slope.

\end{itemize}

In Fig.~\ref{fig:SED} we plot the different components of the SED for a $10\,{\rm M}_{\odot}$ PBH with such fiducial parameters at $z=17$. We see that the SSC and  ICS luminosities are much smaller than the synchrotron. The heating and ionization of the IGM will be dominated by accretion X-ray radiation, if the UV radiation is neglected.

\subsection{ The CRB and 21-cm signal at Cosmic Dawn}

We solve the IGM thermal history and $\dot{m}$  evolution self-consistently, then compute the CRB and the 21-cm signal.
The contributions from accretion X-ray radiation, synchrotron, SSC and the  ICS are all taken into account.

For naked PBHs with $M_{\bullet}=10\,{\rm M}_{\odot}$, we plot $T_{\rm CRB}$, $T_{\rm k}$ and $\delta T_{\rm b}$ at $z=17$ as a function of $f_{\rm PBH}$ for various $f_{\rm DC}$ in Fig.~\ref{fig:temperatures}. For $f_{\rm DC}\lesssim 0.01$, the X-ray heating would be the dominant effect, the 21-cm signal is hardly enhanced. 

In contrast, for  $f_{\rm DC}\gtrsim0.01$, when $f_{\rm PBH} f_{\rm DC}\gtrsim 10^{-5}$, they can significantly contribute to the CRB and boost the 21-cm absorption signal. 
Although the IGM is also heated, $T_{\rm k}$ is still smaller than $T_{\rm CMB}+T_{\rm CRB}$, so the 21-cm signal is still in absorption. For more massive PBHs and a higher abundance, the heating is more efficient, $T_{\rm k}$ can be larger than $T_{\rm CMB}+T_{\rm CRB}$, so it turns 21-cm signal to an emission. However, the resultant high ionization fraction from efficient heating makes the optical depth too large to be compatible with current {\it Planck} constraints. To see this, we calculate the augumented optical depth of the Thomson scattering of CMB photons as
\begin{equation}
\Delta \tau_{\rm e} (>z)= \int^{\infty}_{z} \sigma_{\rm T} n_{\rm H}(z')\Delta x_{\rm e}(z')\frac{c{\rm d}z'}{H(z')(1+z')},      
\end{equation}
where  $\Delta x_{\rm e}$ is the fraction of extra IGM electrons generated by X-rays from PBHs. To avoid distorting the CMB signal, we require that $\Delta \tau_{\rm e}(z>10)$ is smaller than the $2\sigma$ uncertainty level of the {\it Planck} observations, i.e. $\Delta \tau_{\rm e} \lesssim 0.014$~\cite{Planck_1807.06209}.

\begin{figure}
\centering{
\subfigure{\includegraphics[width=0.45\textwidth]{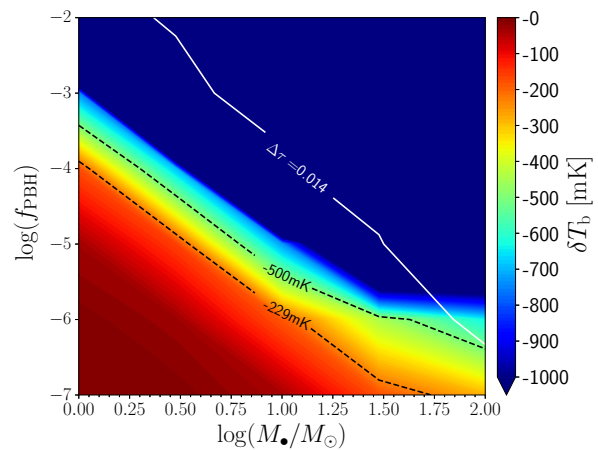}}
\caption{The contour map of $\delta T_{\rm b}$ as a function of PBH mass and abundance, for $f_{\rm DC}=1$. We mark the region where extra CMB optical depth from PBHs at $z>10$ is larger than the {\it Planck} observational $2\sigma$ upper limit (above the solid curve), and the region where the 21-cm absorption can be boosted by the PBHs (above the dashed line).
} 
\label{fig:T_21_contour1}
}
\end{figure}

In Fig.~\ref{fig:T_21_contour1} we plot the contour map of the 21-cm signal with the PBH mass and abundance when $f_{\rm DC}=1$. We see that, even in the presence of X-ray heating, there is still a large parameter space for which the 21-cm absorption can be boosted, 
For example, if $M_{\bullet}\gtrsim 10^2\,{\rm M}_{\odot}$, the 21-cm signal is obviously boosted even $f_{\rm PBH}$ is as low as $\sim10^{-6}$. Even after excluding the parameters for which $\Delta \tau_{\rm e} \ge 0.014$  (above the solid curve), it is still possible to boost the 21-cm absorption signal deeper than $\sim -500$ mK. In Fig. \ref{fig:T_21_contour1} the $\delta T_b = -500$ mK curve corresponds to  $\log(f_{\rm PBH}) = -1.8\log(M_\bullet/{\rm M}_{\odot})-3.5$ for $1\,{\rm M}_{\odot} \lesssim M_\bullet \lesssim 300\,{\rm M}_{\odot}$. For a PBH distribution with a single mass, if $M_{\bullet}\lesssim1\,{\rm M}_{\odot}$ and $f_{\rm PBH} \lesssim10^{-4}$, it will not significantly affect the 21-cm signal.

We also investigate the alternative duty-cycle values, and find that if $f_{\rm DC}\lesssim0.04$, the boosted 21-cm absorption signal would be $\leq 500$ mK if we are to keep $\Delta \tau_{\rm e} \lesssim 0.014$.

\begin{figure}
\centering{
\subfigure{\includegraphics[width=0.45\textwidth]{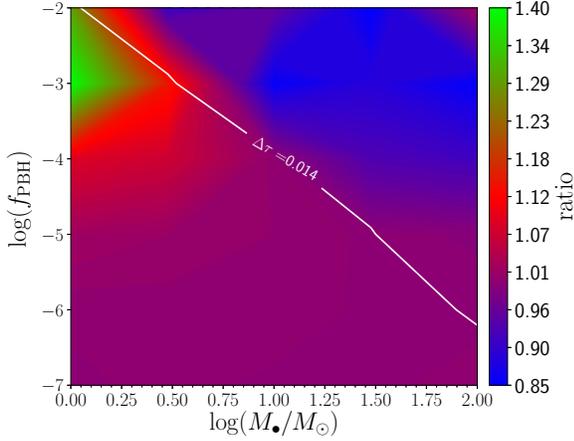}}
\caption{The ratio between the 21 cm signal with and without PBH binaries, jet model parameters are the same as in Fig. \ref{fig:T_21_contour1}. 
} 
\label{fig:T_21_ratio}
}
\end{figure}

In Fig. \ref{fig:T_21_ratio} we show the ratio of the 21 cm signal with PBH binaries to that without PBH binaries, for the fiducial jet model parameters. When $f_{\rm PBH} \lesssim10^{-4}$, the PBH binaries have negligible influence on the 21 cm signal. When $f_{\rm PBH}\gtrsim 10^{-4}$, the PBH binaries boost the 21 cm absorption signal for $M_\bullet \lesssim 10~M_\odot$, while suppress it for $M_\bullet \gtrsim 10~M_\odot$. This is because when the PBHs are heavier, the X-ray heating can somewhat reduce the absorption depth and/or  the accretion rate, by increasing $T_k$, although such parameter region should be ruled out when the reionization constraints are taken into account.  

\begin{figure}
\centering{
\subfigure{\includegraphics[width=0.45\textwidth]{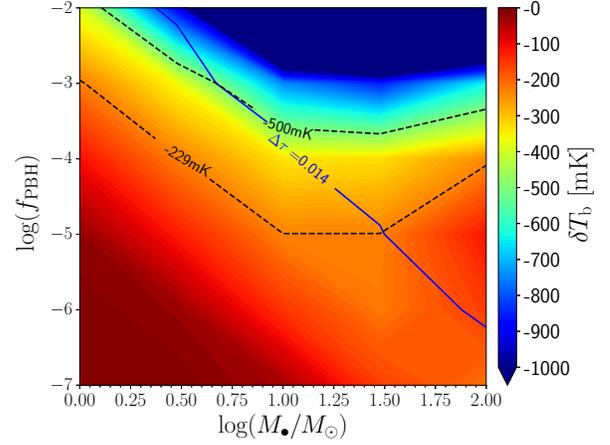}}
\subfigure{\includegraphics[width=0.45\textwidth]{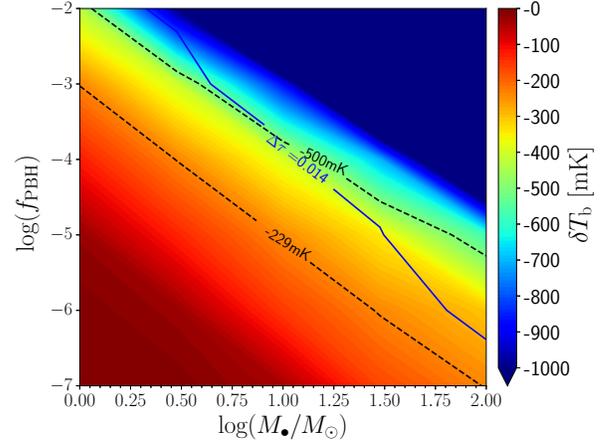}}
\subfigure{\includegraphics[width=0.45\textwidth]{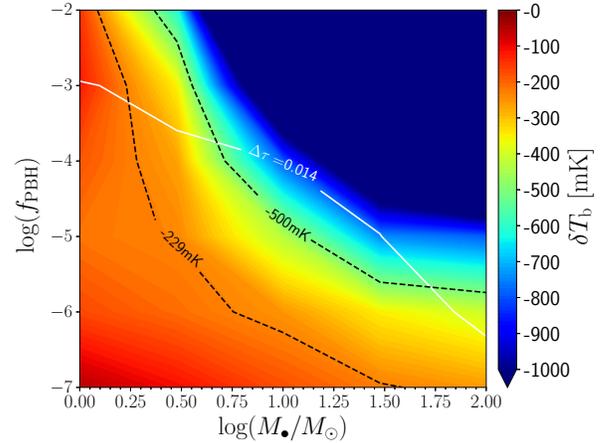}}
\caption{Same to Fig.~\ref{fig:T_21_contour1}, but for $\eta_e\eta_{\rm jet}=0.02$ (top),  or $B=200\,{\rm G}$ (middle), or $s=2.5$ (bottom). 
For each panel, except the specified parameter, other parameters are set to fiducial values. 
}
\label{fig:T_21_contour2}
}
\end{figure}

In Fig.~\ref{fig:T_21_contour2} we show the cases for which either $\eta_e\eta_{\rm jet}$ is lower (top), or $B$ is lower (middle), or the energy distribution of injected relativistic electron is harder (bottom), than our fiducial values respectively. 
 In these cases, we find that if 
$\eta_e\eta_{\rm jet}\lesssim 0.02$, or $B\lesssim 200$ G, or $s\lesssim2.5$, then it would be unable to boost the 21-cm absorption signal to $\sim-500$ mK, while keeping the reionization history almost unchanged at the same time. For the electron distribution with $s=2.5$, the heating by X-rays from the synchrotron is even more important than that from the accretion radiation.

If PBHs are inside dark matter halos, where the gravity of the dark matter halo is still negligible, the accretion rate decreases as the Univese expands, like the naked PBHs. 
However, the rate can increase at a critical redshift because the gravity of the dark matter halo dominates the accretion. This critical redshift depends on the PBH mass, as seen from Fig.~\ref{fig:m_acc_tot}. The 21-cm signal is then more complicated in this case.

In such a model, at very high redshift, when the effects of dark matter halo is still negligible, the 21-cm absorption decreases with decreasing redshift, just as the case for naked PBHs. However, the dark matter halo enhances the PBH gas accretion, so at some point the 21-cm absorption signal starts to increases as redshift decreases, because of the enhancement of CRB. The absorption reaches a maximum  at a redshift $z_{\rm trough}$, then starts to decrease because of the X-ray heating and the expansion of the Universe. Therefore in the presence of dark matter halos, an absorption trough is produced. The depth and redshift of the trough depend on the PBH mass and abundance.

\begin{figure}
\centering{
\subfigure{\includegraphics[width=0.45\textwidth]{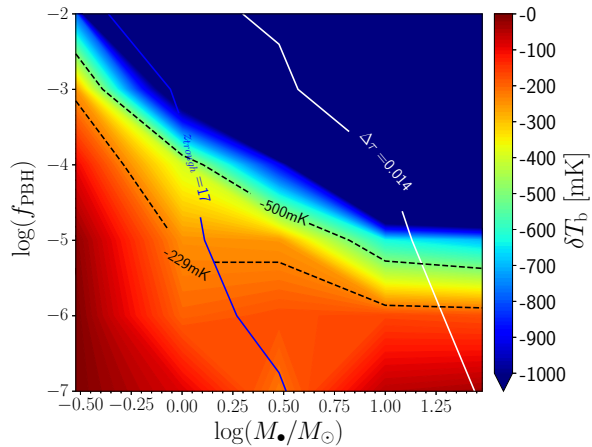}}
\caption{The 21-cm brightness temperature at the absorption trough as a function of $M_{\bullet}$ and $f_{\rm PBH}$, for PBHs inside dark matter halos. We mark the location of $\delta T_{\rm b}(z_{\rm trough})=-500$  mK, and the location of the trough redshift $z_{\rm trough}=17$. The intersection is $M_{\bullet}\sim1.05\,{\rm M}_{\odot}$ and $f_{
\rm PBH}\sim1.5\times10^{-4}$.
} 
\label{fig:T_21_trough}
}
\end{figure}

In Fig.~\ref{fig:T_21_trough} we plot the $\delta T_b$ at $z_{\rm trough}$ as a function of $f_{\rm PBH}$ and $M_\bullet$, assuming $f_{\rm DC}=1$. The $\delta  T_b=-500$ mK curve corresponds to $\log(f_{\rm PBH})= {\rm max}[-1.7\log(M_\bullet/{\rm M}_{\odot})-3.7,-0.2\log(M_\bullet/{\rm M}_{\odot})-5.0]$  for $0.3\,{\rm M}_{\odot} \lesssim M_\bullet \lesssim 30\,{\rm M}_{\odot}$. Since $z_{\rm trough}$ is also a function of $f_{\rm PBH}$ and $M_\bullet$, the cross point of given $\delta T_{\rm b}$ and given $z_{\rm trough}$ curves uniquely determines the mass and abundance of PBHs. 

\begin{figure}
\centering{

\subfigure{\includegraphics[width=0.45\textwidth]{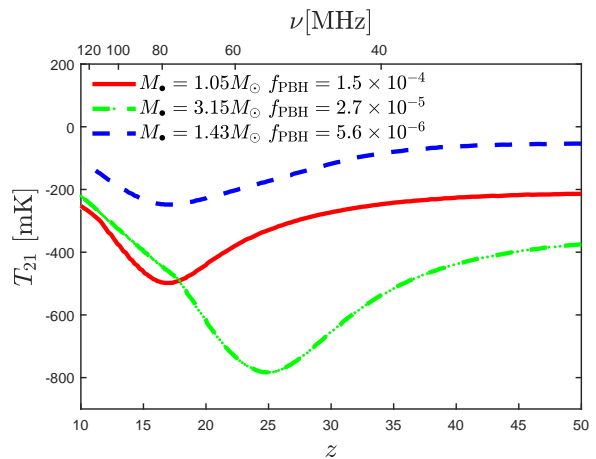}}

\caption{The 21-cm signal as a function of redshift (or  $\nu$) for PBHs inside dark matter halos.
} 
\label{fig:T_21_vs_z}
}
\end{figure}

For example, to produce the maximal 21-cm absorption of $\delta T_{\rm b}=-500$ mK at $z_{\rm trough}=17$, it requires $M_{\bullet}\sim1.05\,{\rm M}_{\odot}$   and $f_{\rm PBH}\sim 1.5\times10^{-4}$. 
In Fig.~\ref{fig:T_21_vs_z} we show $\delta T_{\rm b}$ as a function of redshift for this PBH mass and abundance. For comparison we also plot the curves from some other PBH masses and abundances.  

Similar to the case of naked PBHs, we find that if $f_{\rm DC}\lesssim0.04$ then it would be hard to boost the 21-cm signal significantly.

A dark matter halo does not always boost the accretion rate of its embedded PBH. If a PBH falls into the gravitational well of a previously-formed dark matter halo, then the gravitational potential energy is converted into kinetic energy. In this case, the PBH and ambient medium can have a higher relative velocity than $v_L$. As a result, the accretion is suppressed and the dark matter halo  is not helpful for producing the 21 cm absorption trough. Such a scenario is not involved in our model.

\subsection{The Contribution to Present-day Cosmic Electromagnetic Background}

The redshifted radiation from PBHs at $>z$ forms part of the Cosmic Electromagnetic Background (CEB) in present-day Universe. The intensity can be written as
\begin{equation}
I_{\rm CEB}(\nu,>z)=\frac{1}{4\pi}\int^{\infty}_{z} n_{\bullet} L_{\bullet}(\nu) \frac{c {\rm d}z'}{H(z')(1+z')},    
\end{equation}
where $L_{\bullet}$ is the total PBH luminosity, including the contribution from synchrotron, SSC, ICS of the CMB photons, and the accretion X-ray radiation. For the model to be consistent with observations, we require that the produced background radiation should not exceed what have been observed at different frequency bands. 

\begin{figure}
\centering{
\subfigure{\includegraphics[width=0.45\textwidth]{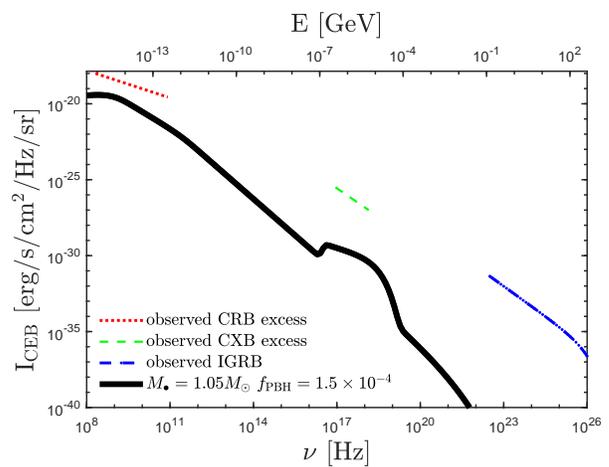}}
\caption{The Cosmic Electromagnetic Background contributed by our model of PBHs inside dark matter halos, assuming $M_{\bullet}=1.05\,{\rm M}_{\odot}$, $f_{\rm PBH}=1.5\times10^{-4}$ and $z>10$. For comparison we plot the observed cosmic radio background excess~\cite{Fixsen2011ApJ}, the observed cosmic X-ray background (CXB) excess~\cite{Cappelluti2017ApJ}, and the observed isotropic gamma-ray background (IGRB)~\cite{Ackermann2015ApJ}. Note that the IGRB should be dominated by the unresolved star-forming galaxies~\cite{Roth2021Natur}. Our PBH contribution is safely smaller than all these observational limits.} 
\label{fig:I_CEB}
}
\end{figure}

In Fig.~\ref{fig:I_CEB} we plot the CEB  from radio to gamma-ray, contributed by PBHs inside dark matter halos at $z>10$, for $M_{\bullet}=1.05\,{\rm M}_{\odot}$  and $f_{\rm PBH}=1.5\times10^{-4}$, 
$f_{\rm DC}=1$.
Compared with the observed CRB excess~\cite{Fixsen2011ApJ}, the cosmic X-ray background excess~\cite{Cappelluti2017ApJ} and the isotropic gamma-ray background (IRGB)~\cite{Ackermann2015ApJ}, the contribution from PBHs in this model is much smaller, so this model is entirely consistent with the existing observations. Of course, it also means that to detect PBHs by using the CEB is a challenge. Such a background must be very isotropic compared with the background built up by astrophysical sources since the dark matter halos hosting the PBHs are much less clustered than galaxies.

\section{Summary  \& Discussion}\label{sec:summary}

In this paper we investigate the contribution to the CRB at Cosmic Dawn from PBHs, and the influence on the 21-cm global spectrum. We assume that a PBH can accrete the ambient IGM, form accretion flow and launch a jet. The radio and X-ray radiation of a PBH is calculated either from an empirical 
$L_R-L_X$ relation derived from observed stellar-mass black holes and supermassive black holes that have jets, or directly from a jet model.

We found the following:

\begin{itemize}

    \item If the empirical $L_R-L_X$ relation is applicable to PBHs,
     generally PBHs can only produce negligible CRB (compared with the CMB) at the Cosmic Dawn, no matter whether they are naked PBHs or inside  dark matter halos. If PBHs produce a stronger CRB (with a larger $M_{\bullet}$  and/or a higher $f_{\rm PBH}$), then they will inevitably heat and ionize the IGM significantly, not only break the observational constraints on the IGM thermal history, but also reduce or even fully erase the 21-cm absorption signal.
     
     However, even when $f_{\rm PBH}$ is small so the heating and ionization of the IGM are negligible,
     PBHs can still produce a strong  Ly$\alpha$ background to couple  $T_{\rm S}$ tightly with $T_{\rm k}$ at Cosmic Dawn. 
     So at least PBHs have the potential to make the 21-cm absorption reach the theoretical maximum in the absence of an extra radiation background, say $\sim-200$ mK.

    \item If the empirical $L_R-L_X$ relation is NOT applicable to PBHs, i.e. the PBHs have strong radio synchrotron radiation from the jets, without producing as much X-ray as inferred from the observed $L_R-L_X$ relation, they can produce a CRB that is strong enough to boost the 21-cm absorption signal to $\sim -500$ mK at the Cosmic Dawn, while at the same time only cause negligible or modest heating/ionization on the IGM, as long as
$\log(f_{\rm PBH}) = -1.8\log(M_\bullet)-3.5$ for $1\,{\rm M}_{\odot} \lesssim M_\bullet \lesssim 300\,{\rm M}_{\odot}$.
    
    In this case, the required minimum fraction of power finally converted into relativistic electrons compared with the accretion power is $\eta_e\eta_{\rm jet}\sim0.02$;
    or minimum magnetic field is $\sim 200$ G, or the minimum spectrum index of injected relativistic electron distribution is $\sim 2.5$.  These parameter values are quite plausible, at least for the observed stellar-mass black holes with jets.
    
    \item If PBHs are inside the dark matter halos, at first the accretion rate decreases with decreasing redshift, but then starts to increase since the dark matter halos help attracting the gas.  As a result, the 21-cm absorption reaches a maximum at some redshift $z_{\rm trough}$. 
    Before this time both the CRB and heating are weak, so the 21-cm absorption is weak; after this time, the heating is more efficient so that the 21-cm absorption is reduced even though the CRB keeps increasing, producing a trough in the 21-cm global spectrum $\delta T_{\rm b}(\nu)$. The redshift and depth of the the trough depends on the PBH mass and abundance. For $M_{\bullet}=1.05\,{\rm M}_{\odot}$ and $f_{\rm PBH}=1.5\times10^{-4}$, it gives  $z_{\rm trough}\sim 17$ and $\delta T_{\rm b}(z_{\rm trough})\sim -500$ mK.

\end{itemize}

In our model, for PBHs inside dark matter halos, if $M_{\bullet}=30\,{\rm M}_{\odot}$ and $z=10$, the radio flux of one such PBH at $1.4$ GHz is less than $10^{-5}$ mJy.  This is far below the SKA sensitivity \citep{Braun2019_SKA} limit, so  individual emission of such objects are still beyond the reach of the upcoming large radio telescope. The influence on the 21-cm signal provides some potential for revealing its properties.

Launching a jet requires the angular momentum supply. Suppose that the angular momentum of a PBH gradually grows via accretion, then the PBH can only launch a jet after a sufficiently long time of accretion which builds up its angular momentum. Ref. \cite{2003.02778} provided an intensive study about the growth of spin for PBHs with various masses. They found that: for the PBHs smaller than $\mathcal{O}(30)~M_\odot$, the spin is likely always negligible. However, for heavier PBHs, the spin increases fast via super-Eddington accretion  after $z\lesssim30$, and finally reaches the maximum in timescale $\lesssim$ Hubble time. It provides not only the necessary mechanism for a PBH to launch the jet, but also a natural solution to the problem of PBHs generate too strong 21-cm absorption signal at stage much earlier than  Cosmic Dawn in the naked PBH models. However, a quantitative investigation is beyond the scope of this paper, we leave it to future work.

For a regular black hole, the jet is supplied by its spin, 
and the spindown timescale is $\sim0.2$ Gyr~\cite{jet_2108.12380}, which is comparable to the Hubble time at $z=17$.   Therefore the assumption for a $f_{\rm DC}\sim\mathcal{O}(1)$ is not unreasonable.  
However, it is possible that some PBHs may not have enough time to collect sufficient angular momentum to launch the jet, so $f_{\rm DC}$ can also be significantly smaller than 1. In this case, our predicted 21-cm absorption should be reduced by a factor  $\sim f_{\rm DC}$. In other words, to produce the same amount of the signal, the required PBH abundance should be boosted by a factor $\sim 1/f_{\rm DC}$. But we do not expect $f_{\rm DC}\lesssim0.01$, otherwise either the required $f_{\rm PBH}$ is too high, or the X-ray heating effect will erase the 21-cm absorption.

We also estimated the contributions from PBH binaries to X-ray heating and CRB. If the PBHs are highly clustered, for example $\delta_{\rm dc}=1000$, then their binaries may have important contributions when $f_{\rm PBH}\gtrsim 10^{-4}-10^{-3}$. 
Such an estimation is preliminary, but could be considered as a useful references,
though the following ignored effects should be noted:

a) For dense star/PBH clusters, if the accretion radii of the members overlap significantly, then the clusters will be in the coherent accretion mode, for which the accretion rate of each member is boosted by a factor of $N_{\rm c}$, where $N_{\rm c}$ is the number of cluster members \cite{Lin2007ApJ,Kaaz2019ApJ,Hutsi2019PhRvD}. Ref. \cite{Hutsi2019PhRvD} pointed out that for PBHs the coherent boost effect is sizable when $M_\bullet \gtrsim 10~{\rm M}_\odot$ and $f_{\rm PBH}\gtrsim 0.1$.

b) PBH binaries formed much earlier than the Cosmic Dawn. Those with small semi-major axes have short coalescence time, and can therefore merge into single black holes before the Cosmic Dawn \cite{Peters1964PhRv}. Moreover, the merger rate is generally enhanced if PBH binaries are located in clusters \cite{Raidal2017JCAP,Young2020JCAP,DeLuca2020JCAP}. Mergers may modify the role played by PBH binaries. However, PBH binaries can be either disrupted   or hardened (the binding energy is increased) when they interact with the surrounding PBHs \cite{ Raidal2017JCAP,Raidal2019JCAP,Young2020JCAP,DeLuca2020JCAP,Jedamzik2020JCAP,Vaskonen2020PhRvD}, and the interaction may happen frequently in clusters. Therefore the merger rate and the role played by the PBH binaries have 
 complicated dependence on the clustering and $f_{\rm PBH}$. For example, Ref. \cite{Atal2020JCAP} found that when $f_{\rm PBH}\gtrsim 
 10^{-3}$, in clusters the merger rate decreases  with increasing $f_{\rm  PBH}$.

There are observational constraints on the abundance of PBHs based on its non-radiative characteristics. For example, CMB anisotropy implies $f_{\rm PBH} \lesssim 10^{-1}-10^{-6}$ for PBH masses from $M_{\bullet}=1\,{\rm M}_{\odot}$ to  $M_{\bullet}=10^2\,{\rm M}_{\odot}$~\cite{CMBbounds_Serpico_2002.10771}; the constraints from stochastic background of gravitational wave possibly produced by PBH mergers gives $f_{\rm PBH} \lesssim 10^{-2}-10^{-3}$ for $M_{\bullet}=1\,{\rm M}_{\odot}$ to $10^2\,{\rm M}_{\odot}$~\cite{1904.02396}; and gravitational lensing imposes a bound of $f_{\rm PBH} 
\lesssim 10^{-1}$ in stellar mass range~\cite{lensing_PBH_1710.00148}.  The PBH abundance and mass for producing the strong 21-cm absorption signal as considered in this work are consistent with these constraints.

Ref. \cite{Hasinger2020JCAP} found that the PBHs can easily produce sufficient CRB to explain the  21 cm absorption depth as observed by EDGES, provided that $\sim$5\% of them have radio luminosity 1000 times higher than the value predicted by the empirical $L_R-L_X$ relation in Ref.  \cite{Wang2006ApJ}. It implies that, even ignoring the X-ray heating, at $z\sim 17$ the $T_{
\rm CRB}$ from the PBHs must be $\gtrsim 2T_{\rm CMB}$. 
Our empirical model adopted the $L_R-L_X$ relation for black holes with jet activities in Ref. \cite{Merloni2003}, and incorporated the X-ray heating/ionization and their feedback self-consistently. The results of our empirical model are not as optimistic as Ref. \cite{Hasinger2020JCAP}, probably because Ref. \cite{Hasinger2020JCAP} adopted a complicated PBH mass spectrum and allowed the cumulative $f_{\rm PBH}\sim1$, while we investigated the case with monochromatic mass in the limit of $f_{\rm PBH}\ll 1$.  
We further investigated the PBH radio signal in greater detail by using a jet model, and found the jet parameters required for the PBHs to produce the amount of CRB that can explain the EDGES results. We found that it requires a strong magnetic field and a high fraction of energy deposited into relativistic electrons, which is still physically feasible.

\begin{acknowledgments}
We thank Dr. Erlin Qiao for the very helpful discussions.
This work is supported by National SKA Program of China no. 2020SKA0110402 and no. 2020SKA0110401, the Inter-Governmental Cooperation China-South Africa Program grant 2018YFE0120800, the Chinese Academy of Science (CAS) Scientific Research Instrument and Equipment Development Project ZDKYYQ20200008, the National Natural Science Foundation of China grant No. 11973047, and CAS frontier research grant QYZDJ-SSW-SLH017. Y.Z.M. is supported by the National Research Foundation of South Africa under grant No. 120385 and No. 120378, NITheCS program ``New Insights into Astrophysics and Cosmology with Theoretical Models confronting Observational Data'', and National Natural Science Foundation of China with project 12047503. This work used computing resources of the Astronomical Big Data Joint Research Center, co-founded by National Astronomical Observatories, Chinese Academy of Sciences and Alibaba Cloud.

\end{acknowledgments}



\bibliography{ms}

\end{document}